\documentclass[draft]{agujournal2019}
\usepackage{url} 
\usepackage{lineno}
\usepackage[inline]{trackchanges} 
\usepackage{soul}
\usepackage{physics}

\draftfalse

\journalname{JGR: Planets}

\begin{document}

%
%


\title{Latitudinal variations in methane abundance, aerosol opacity and aerosol scattering efficiency in Neptune's atmosphere determined from VLT/MUSE}

%
%




\authors{P. G. J. Irwin\affil{1}, J. Dobinson\affil{1}, A. James\affil{1}, M. H. Wong\affil{2}, L. N. Fletcher\affil{3}, M. T. Roman\affil{3}, N.A. Teanby\affil{4}, D. Toledo\affil{5}, G.S. Orton\affil{6}, S. P\'{e}rez-Hoyos\affil{7}, A. S\'{a}nchez-Lavega\affil{7}, A. Simon\affil{8}, R. Morales-Juberias\affil{9}, I. de Pater\affil{10}}


\affiliation{1}{Atmospheric, Oceanic and Planetary Physics, Department of Physics, University of Oxford, Parks Rd, Oxford, OX1 3PU, UK}
\affiliation{2}{Center for Integrative Planetary Science, University of California, Berkeley, CA 94720-3411, USA}
\affiliation{3}{School of Physics \& Astronomy, University of Leicester, University Road, Leicester, LE1 7RH, UK}
\affiliation{4}{School of Earth Sciences, University of Bristol, Wills Memorial Building, Queens Road, Bristol, BS8~1RJ, UK}
\affiliation{5}{Instituto Nacional de T\'ecnica Aeroespacial (INTA), 28850, Torrej\'on de Ardoz (Madrid), Spain.}
\affiliation{6}{Jet Propulsion Laboratory, California Institute of Technology, 4800 Oak Grove Drive, Pasadena, CA~91109, USA}
\affiliation{7}{University of the Basque Country UPV/EHU, 48013 Bilbao, Spain}
\affiliation{8}{Solar System Exploration Division/690, NASA Goddard Space Flight Center, 8800 Greenbelt Rd, Greenbelt, MA~20771, USA}
\affiliation{9}{New Mexico Institute of Technology, Soccoro, New Mexico, USA}
\affiliation{10}{Department of Astronomy and Department of Earth and Planetary Science, University of California, Berkeley, CA 94720, USA}





\correspondingauthor{Patrick Irwin}{patrick.irwin@physics.ox.ac.uk}




\begin{keypoints}
\item Neptune MUSE visible/near-infrared spectra are well fitted by a simple aerosol model comprised of three distinct layers
\item A darkening of particles at blue-green wavelengths in a deep aerosol layer near 5 bar can explain dark spots and the dark `South Polar Wave'
\item A brightening of the same particles at red-infrared wavelengths can explain bright zones and spots seen in longwave narrow reflectance peaks

\end{keypoints}

%
%

%
%


\begin{abstract}
Spectral observations of Neptune made in 2019 with the MUSE instrument at the Very Large Telescope in Chile have been analysed to determine the spatial variation of aerosol scattering properties and methane abundance in Neptune's atmosphere. The darkening of the South Polar Wave (SPW) at $\sim$ 60$^\circ$S, and dark spots such as the Voyager 2 Great Dark Spot is concluded to be due to a spectrally-dependent darkening ($\lambda < 650$nm) of particles in a deep aerosol layer at $\sim$ 5 bar and presumed to be composed of a mixture of photochemically-generated haze and H$_2$S ice. We also note a regular latitudinal variation of reflectivity at wavelengths of very low methane absorption longer than $\sim$ 650 nm, with bright zones latitudinally separated by $\sim$ 25$^\circ$. This feature, similar to the spectral characteristics of a discrete deep bright spot DBS-2019 found in our data, is found to be consistent with a brightening of the particles in the same $\sim$5-bar aerosol layer at $\lambda > 650 $ nm. We find the properties of an overlying methane/haze aerosol layer at $\sim$ 2 bar are, to first-order, invariant with latitude, while variations in the opacity of an upper tropospheric haze layer reproduce the observed reflectivity at methane-absorbing wavelengths, with higher abundances found at the equator and also in a narrow `zone' at $80^\circ$S. Finally, we find the mean abundance of methane below its condensation level to be 6--7\% at the equator reducing to $\sim$3\% south of $\sim$25$^\circ$S, although the absolute abundances are model dependent.
\end{abstract}

\section*{Plain Language Summary}
Observations of Neptune in visible light, made with the MUSE instrument at ESO’s Very Large Telescope,  reveal the different layers of clouds and gases within this Ice Giant atmosphere, and how they change with height and latitude.  The properties of the 1-2-bar layer of methane ice and haze are found to be roughly constant with latitude. However,  a diffuse upper layer is thickest at the equator and near the south pole, indicating air rising at mid-latitudes and descending near the equator and poles.  Conversely, the distribution of methane between the deep 5-bar clouds (hydrogen sulphide ice and haze) and the middle layers decreased from 6--7\% at the equator to $\sim$3\% near the south pole, suggesting rising air instead at the equator and descending elsewhere. Finally, a blue-green darkening of the particles in the deep layer can explain Neptune's dark spots and the dark `South Polar Wave' at 60$^\circ$S, whereas a brightening of the same particles at red and infrared wavelengths matches occasional discrete deep bright spots and a sequence of previously unnoticed bright `zones’, separated by ~25$^\circ$ latitude. All this is evidence that the atmospheric circulation changes as a function of height and latitude in complex and surprising ways.

\section{Introduction} \label{introduction}

The study of the aerosols in Neptune's atmosphere has been transformed in recent years by the analysis of visible and near-infrared multi-spectral imaging data. The retrieval of vertical profiles of aerosols from such data requires knowledge of the vertical and latitudinal distribution of the main gaseous absorber, methane, which for many years, in the absence of any other information, was assumed to be constant at all latitudes. However, by analysing Hubble Space Telescope (HST) Space Telescope Imaging Spectrograph (STIS) observations of Neptune from 2003, \citeA{kark11} showed that the abundance of methane at equatorial latitudes ($\sim4$\%) was approximately twice that detected at polar latitudes ($\sim2$\%), a result that had significant repercussions on the inferred vertical structure of aerosols, which were subsequently found to vary less significantly with latitude. In 2018, Neptune was observed during commissioning operations with the Multi Unit Spectroscopic Explorer (MUSE) Integral Field Unit (IFU) Spectrometer at the Very Large Telescope (VLT) at the European Southern Observatory in Chile. An initial analysis of these data \cite{irwin19methane} found a similar latitudinal variation of cloud-top methane mole fraction as the HST/STIS study, and was later revised to include modelling of the limb-darkening effects \cite{irwin21} using the Minnaert approximation \cite{minnaert41}. \citeA{irwin21} concluded that the `deep' (i.e., at 2--4 bar) mole fraction of methane varied from 4--6\% at the equator to 2--4\% at polar latitudes, with a boundary at $\sim$30$^\circ$S. Most recently, a joint analysis of HST/STIS, Gemini/NIFS and IRTF/SpeX observations of Neptune and Uranus from 0.3 to 2.5 $\mu$m \cite{irwin22} has developed a `holistic' aerosol model of both planets comprised of three basic distinct aerosol layers:  1) a deep H$_2$S/photochemical-haze aerosol layer with a base pressure $\ge$ 5--7 bar (Aerosol-1); 2) a layer of methane/photochemical-haze just above the methane condensation level at 1--2 bar (Aerosol-2); and 3) an extended layer of small photochemical haze particles extending into the stratosphere (Aerosol-3). For Neptune an additional contribution from upper level ($\sim 0.2$ bar) methane ice clouds was required to match the IRTF/SpeX observations at $\lambda >$ 1 $\mu$m.

The atmosphere of Neptune occasionally displays dark spots, seen at blue-green  visible wavelengths $<$ 650 nm, including the Great Dark Spot (GDS) observed by Voyager 2 in 1989 \cite{smith89}, and more recent examples captured in HST Wide Field Camera 3 (WFC3) observations \cite<e.g.,>{hueso19}. The most recent dark spot was discovered by HST/WFC3 in 2018 at 23$^\circ$N and named NDS-2018  \cite{simon19}. NDS-2018 was of a similar size to the GDS and subsequently drifted equatorwards, disappearing in 2022 \cite{wong22b}. As part of a global effort to observe and analyse NDS-2018, Neptune was observed with VLT/MUSE in late 2019 \cite{irwin23}. After spatial deconvolution, \citeA{irwin23} were able to detect the NDS-2018 spot at $\sim$$15^\circ$N in these observations (the first ground-based detection of such a spot), and were also able to spectrally characterise it at high spectral resolution, the first time that this has ever been achieved for a Neptunian dark spot. \citeA{irwin23} found that NDS-2018 was caused by a spectrally-dependent ($\lambda <$ 650 nm) darkening of the particles in the Aerosol-1 layer at $\sim$5 bar and also found a nearby deep bright spot, DBS-2019 at $\sim 10^\circ$N, which they concluded was caused by a brightening of the same layer at longer wavelengths. 

In addition to capturing the reflection spectra of NDS-2018 and DBS-2019, the 2019 VLT/MUSE data also show distinct latitudinal variations over a wide range of wavelengths (473 -- 933 nm), in particular the South Polar Wave (SPW) at $\sim$60~$^\circ$S, first seen in Voyager 2 images in 1989 \cite{smith89}, which is dark at blue-green wavelengths, but invisible at longer wavelengths. The spectral features of the SPW are very similar to those of dark spots and \citeA{karkoschka11_dark} concluded that it is caused by a darkening of particles at pressures $>$ 3 bar, which was later confirmed by \citeA{irwin22}. The SPW has been visible ever since the Voyager 2 flyby in HST imaging observations and these observations are reviewed in detail by \citeA{karkoschka11_rotation}. The SPW is found to have a southern boundary at $\sim70^\circ$S, and a clearer northern boundary that varies with longitude as a wavenumber-1 disturbance of amplitude $\sim 5^\circ$, and whose northern extent varies with time from 50 -- 55$^\circ$S. The SPW feature seems to be related to the South Polar Feature (SPF) near 70$^\circ$S, which are short-lived, bright clouds that change on a time scale of hours \cite{hammel89}. \citeA{karkoschka11_rotation} notes that the SPF longitude coincides with that of the northernmost extent of the SPW, which suggests there is a dynamical link. The features may also have been dynamically linked with the second Voyager dark spot (DS2) \cite{sromovsky93} and \citeA{karkoschka11_rotation} suggest that these features are static with respect to a new coordinate system for the deep rotation of Neptune with a rotation rate of 15.9663 hours, compared with 16.108 hours derived from Voyager radio data \cite{lecacheux93}. 

While the 2019 VLT/MUSE data show the SPW very clearly, they also detected latitudinal variations at several other wavelengths that have not been fully noted before. In this paper, we further analyse the 2019 VLT/MUSE data  to revise our understanding of the latitudinal variation in aerosol properties and methane abundance in Neptune's atmosphere.

\section{Observations} \label{observations}

The observations of Neptune were recorded in October and November 2019 using the Multi Unit Spectroscopic Explorer (MUSE) Integral Field Unit (IFU) spectrometer \cite{bacon10}  at the Very Large Telescope (VLT) of the European Southern Observatory at La Paranal, Chile. MUSE records `cubes' of data, where each point in the 300 $\times$ 300 `spaxel' image contains a complete spectrum covering $\sim$ 3700 wavelengths from 473 to 933 nm at a spectral resolution of 2000 -- 4000. The observations were recorded in Narrow-Field Mode, which has field of view of 7.5" $\times$ 7.5" and each `spaxel' is of size 0.025" $\times$ 0.025" (equivalent to 530 km $\times$ 530 km on Neptune's disc). To improve the signal-to-noise ratio, and also to make the observations consistent with our only available source of methane absorption data \cite{kark10}, the spectra were averaged with a triangular instrument lineshape of Full-Width-Half-Maximum (FWHM) 2 nm, sampled every 1 nm  (to achieve Nyquist sampling), reducing the number of wavelengths from $\sim$ 3700 to 459.

\begin{table}[!h]
\setlength{\tabcolsep}{2pt}
\caption{VLT/MUSE Neptune observations (Narrow-Field Mode).\label{tbl-1}}
\begin {tabular}{l l l l l l }
\hline

Obs. & Date & Time & airmass & seeing   &Observing Conditions \\
ID   &      & (UT) &         & (arcsec) &(and any features present)\\
\hline
1A & Oct. 17th 2019 & 00:10:35 & 1.208 & 0.88 & Poor\\
1B & Oct. 17th 2019 & 00:15:20 & 1.194 & 0.84 & Poor, incomplete\\
1C & Oct. 17th 2019 & 00:20:09 & 1.181 & 0.90 & Poor\\
1 & Oct. 17th 2019 & 00:24:51 & 1.170 & 0.69 & Poor\\
2 & Oct. 17th 2019 & 03:47:29 & 1.150 & 0.63 & Moderate\\
3 & Oct. 17th 2019 & 03:52:12 & 1.160 & 0.74 & Moderate\\
4 & Oct. 17th 2019 & 03:56:55 & 1.171 & 0.76 & Moderate\\
5 & Oct. 17th 2019 & 04:01:38 & 1.183 & 0.69 & Moderate\\
\textbf{6} & \textbf{Oct. 18th 2019} & \textbf{00:01:20} & \textbf{1.223} & \textbf{0.67} & \textbf{Clear, NDS-2018 and DBS-2019}\\
7 & Oct. 18th 2019 & 00:18:08 & 1.176 & 0.62 & Clear, NDS-2018 and DBS-2019\\
8 & Oct. 18th 2019 & 00:22:53 & 1.165 & 0.58 & Clear, NDS-2018 and DBS-2019\\
9 & Oct. 18th 2019 & 00:27:36 & 1.154 & 0.55 & Clear, NDS-2018 and DBS-2019\\
10 & Oct. 18th 2019 & 00:32:19 & 1.144 & 0.56 & Clear, NDS-2018 and DBS-2019\\
11 & Oct. 18th 2019 & 03:26:42 & 1.118 & 0.62 & Clear/Moderate\\
12 & Oct. 18th 2019 & 03:32:01 & 1.127 & 0.76 & Clear/Moderate\\
13 & Oct. 18th 2019 & 03:37:33 & 1.138 & 0.70 & Clear/Moderate\\
14 & Oct. 18th 2019 & 03:42:53 & 1.149 & 0.70 & Clear/Moderate\\
15 & Nov. 13th 2019 & 02:31:22 & 1.232 & 0.93 & Poor\\
16 & Nov. 13th 2019 & 02:36:05 & 1.248 & 1.00 & Poor/Moderate\\
17 & Nov. 13th 2019 & 02:40:48 & 1.264 & 0.97 & Poor/Moderate\\
18 & Nov. 13th 2019 & 02:35:33 & 1.281 & 0.99 & Poor/Moderate\\
\hline
\multicolumn{6}{@{}p{\textwidth}@{}}{\footnotesize N.B., the Observation ID mostly relates to order of observation in our data set. Exposure time for all observations was 120s. Obs `6' is highlighted as this was the main set used in this study.} 
\end {tabular}
\end{table}

\begin{figure*}[h]%
\centering
\includegraphics[width=1.0\textwidth]{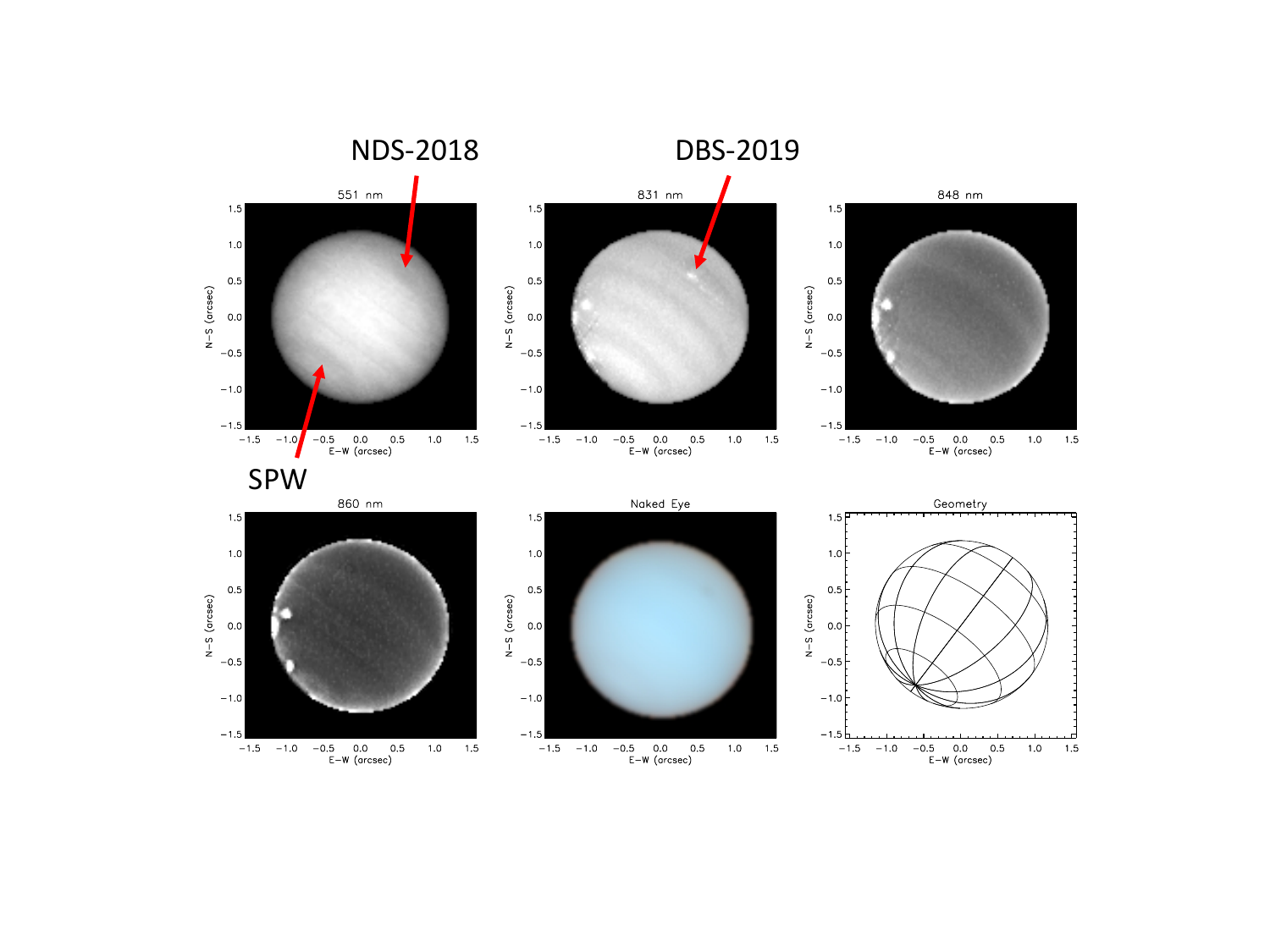}
\caption{Deconvolved MUSE `slices' at 551, 831, 848 nm, and 860 nm, respectively, from the `Obs-6' cube. Also shown is the modelled appearance of Neptune to the naked eye, reconstructed from the MUSE observations using standard colour-matching functions and correctly accounting for gamma corrections. A reference latitude and longitude grid, with spacing of 30$^\circ$ in each direction, is also shown. In the 511-nm and naked-eye images the South Polar Wave (SPW) is just visible at bottom left, while the dark spot NDS-2018 can just be seen at top right. The DBS-2019 bright spot is only visible at longer wavelengths of extremely low methane absorption, such as 831 nm shown here. The 848 and 860 nm images, at methane-absorbing wavelengths, probe the haze high in the atmosphere, with 860 nm probing slightly higher and revealing a south polar collar at 80$^\circ$S. The bright spots on the left edge of Neptune's disc at longer wavelengths are upper tropospheric methane ice clouds at 0.1 - 0.6 bar.}\label{museimages}
\end{figure*}

Neptune was observed on several occasions in this programme (summarised in Table \ref{tbl-1}), where five exposures on October 18th (Observations 6 -- 10), included the NDS-2018 dark spot. 
Although these data were recorded with the GALACSI adaptive optics system \cite{stuik06}, using a laser guide star, the achieved spatial resolution near 500 nm was insufficient to resolve the faint NDS-2018 feature, which has a diameter of $\sim$0.2" and has low contrast. However, with the development of a novel deconvolution technique, \textsc{modified-clean}, the deconvolved `cubes' had sufficient discrimination to detect and spectrally characterise the NDS-2018 dark spot and also a nearby deep bright spot DBS-2019 \cite{irwin23}. The appearance of Neptune at several wavelengths in the best-resolved deconvolved cube in this set, `Obs-6' is shown in Fig. \ref{museimages}. \citeA{irwin23} found that the NDS-2018 dark spot (visible at 551 nm in Fig. \ref{museimages}) is formed by a spectrally-dependent ($\lambda < $ 650 nm) darkening of the aerosols in the deep Aerosol-1 layer at $\sim$5 bar, presumably coincident with the H$_2$S condensation level. \citeA{irwin23} also found a new deep bright spot near NDS-2018, visible in the 831-nm image, which they named `Deep Bright Spot - 2019' (DBS-2019), and which, from the narrowness of its reflectance peaks, was determined to be caused by a spectrally-dependent brightening of the same $\sim$5-bar Aerosol-1 layer at wavelengths longer than $\sim$ 650 nm.

While \citeA{irwin23} describe the characterisation and interpretation of the NDS-2018 and DBS-2019 discrete features, the MUSE data also have sufficient spatial and spectral resolution to determine latitudinal variations of aerosol opacity and methane abundance. These latitudinal changes are clearly visible in Fig. \ref{museimages}, including the dark SPW at $60^\circ$S at short wavelengths (here at 551 nm), but also prominent banded features at longer continuum wavelengths (here at 831 nm), and a small bright collar (80$^\circ$S) seen about the south pole at 860 nm, which at a wavelength of strong methane absorption must be caused by increased aerosol abundance high in the atmosphere. The banding seen at 831 nm is also just visible in recent 845-nm HST/WFC3 images \cite<e.g.,>{chavez23b}, which covers the same reflectance peak, but at much lower spectral resolution ($\Delta \lambda \sim 84$ nm). However, the banding is particularly prominent at this wavelength, which is at the centre of reflectance peak where we can see to deep pressures in the atmosphere, and the narrowness of the spectral signature of these bands, discussed later, indicate that they are likely caused by variations of the deep aerosol layers. 

Although the MUSE observations were photometrically corrected by observing a standard star shortly before viewing Neptune, uncertainty remained in the absolute correction. Hence, to ensure the disc-averaged MUSE spectrum was consistent with the disc-averaged spectrum of Neptune measured in 2003 with HST/STIS \cite{kark11}, the data were scaled to give the same reflectivity as HST/STIS in the equatorial region of 10$^\circ$S -- 10$^\circ$N. Long-term records of the disc-integrated blue and green magnitudes of Neptune \cite{lockwood19} reveal no notable changes between 2003 and the 2016, which justifies this simple scaling and which also has the merit of allowing direct comparison with retrievals from the HST/STIS observations.

\section{Analysis} \label{analysis}
The analysis in this paper follows on from a combined analysis of HST/STIS, IRTF/SpeX, and Gemini/NIFS observations of both Uranus and Neptune by \citeA{irwin22}, which we will refer to henceforth as `IRW22' or the `holistic' model. Given that we do not know the composition of the aerosols in Neptune's atmosphere, and do not have a definite expectation of the volume mixing ratio profile of the main visible/IR absorber, methane, the simultaneous retrieval of aerosol and methane abundance profiles is a degenerate problem, even when analysing the 800 -- 860 nm region that can differentiate between aerosols and methane abundance \cite{kark11}. As a result, since previous studies have concentrated on particular narrow wavelength ranges, this has led to a multitude of similar, but not wholly consistent aerosol/methane solutions. One way to reduce the degeneracy is to analyse simultaneously as wide a wavelength range as possible in order that the sizes of the scattering particles can be better constrained (small particles will have cross-sections that fall as $1/\lambda^4$, while larger particles will have cross-sections that vary more slowly with wavelength). The wavelength range analysed by IRW22 was 0.3 -- 2.4 $\mu$m, which allowed good discrimination against particle size. However, even when analysing a wide range of wavelengths, there are still multiple solutions as we do not know the composition of particles and so we must try to infer the scattering properties (in this case the opacities and single-scattering albedos) directly from the data. IRW22 addressed this degeneracy by attempting to fit simultaneously not only the disc-averaged reflectance  spectrum over this range, but also how the spectrum varied with solar and viewing zenith angles, through fitting their `limb-darkening' or `centre-to-limb' functions. This was done by analysing all observations on the planets' discs and quantifying how the reflectance at each wavelength, $(I/F)$, varied towards the limb of the planet. IRW22 found that this variation was well approximated by the Minnaert approximation \cite{minnaert41}:

\begin{equation}\label{eq:minnaert}
(I/F) = (I/F)_{0} \mu_{0}^{k} \mu^{k-1}.
\end{equation}

Here, $\mu$ and $\mu_0$ are the cosines of the viewing and solar zenith angles, respectively, $(I/F)_0$ is the fitted nadir reflectance, and $k$ the fitted limb-darkening parameter. The reflectance, $I/F$, is the observed radiance at each location, $I$, divided by the radiance reflected from a perfectly-reflecting Lambertian surface at the same point, and in this study we used the solar spectrum of \citeA{chance10}.  IRW22 estimated the scattering properties of the particles in the different aerosol layers using Mie theory and fitted for their complex refractive index spectra. Note that this analysis assumes that the particles in a certain layer, which might be a mixture of ice and haze particles of various size distributions, can be approximated as being composed of a single size distribution of particles, with a single complex refractive index spectrum. The imaginary refractive index spectra were treated as free parameters to be fitted, while the real parts were computed using a Kramers-Kronig analysis, assuming $n_{real} = 1.4$ at 800 nm; this value of $n_{real}$ is typical of giant planet condensates, such as methane ice, but is not well constrained (other values were tested and gave similar overall results). We note that the Kramers-Kronig approach requires that we know $n_{imag}(\lambda)$ at all wavelengths to accurately reproduce $n_{real}(\lambda)$, whereas in this study $n_{imag}(\lambda)$ could only be estimated from 0.3 to 2.4 $\mu$m. Hence, this reconstruction of $n_{real}(\lambda)$ is only an approximation. Mie theory was then  used to compute the extinction cross-section and single-scattering albedo of the aerosols as a function of wavelength. Finally, the Mie-calculated phase functions were approximated with combined Henyey-Greenstein phase functions at each wavelength to average over features particular to spherical particles, namely the `rainbow' peak and the back-scattered `glory'. Although containing several approximations, this approach, first presented by \citeA{irwin15}, has the merit of generating self-consistent values of the extinction cross-section, single-scattering albedo, and phase function spectra, and has considerable advantages, we believe, over approaches where the single-scattering albedo (or phase functions, or extinction coefficients) are modified directly and separately. The holistic IRW22 analysis  found that the combined HST/STIS, IRTF/SpeX and Gemini/NIFS observations of both Uranus and Neptune were well approximated with an atmospheric model consisting of three main aerosol layers, outlined earlier:  1) `Aerosol-1', a deep aerosol layer with a base pressure $\ge$ 5--7 bar, assumed to be composed of a mixture of H$_2$S ice and photochemical haze of mean radius 0.05 $\mu$m; 2) `Aerosol-2', a layer of photochemical haze and methane ice of mean radius $\sim 0.5$ $\mu$m, confined within  a layer of high static stability at the methane condensation level at 1--2 bar; and 3) `Aerosol-3', an extended layer of small photochemical haze particles ($r \sim 0.05$ $\mu$m), probably of the same composition as the haze in the 1--2-bar layer, extending from this level up through to the stratosphere.  An additional thin layer of micron-sized ($r \sim$ 3 $\mu$m) methane ice particles at $\sim$0.2 bar (`Aerosol-4') was required to explain enhanced reflection at methane-absorbing wavelengths in the Neptune data longer than 1.0 $\mu$m, caused probably by discrete clouds at this level being present along the central meridian of the IRTF/SpeX line-averaged data used, and unresolved clouds in the Gemini/NIFS observations. The region of high static stability at the methane condensation level has been noted by several previous authors \cite<e.g.,>{hueso20,leconte17}.

\begin{figure*}[h]%
\centering
\includegraphics[width=1.0\textwidth]{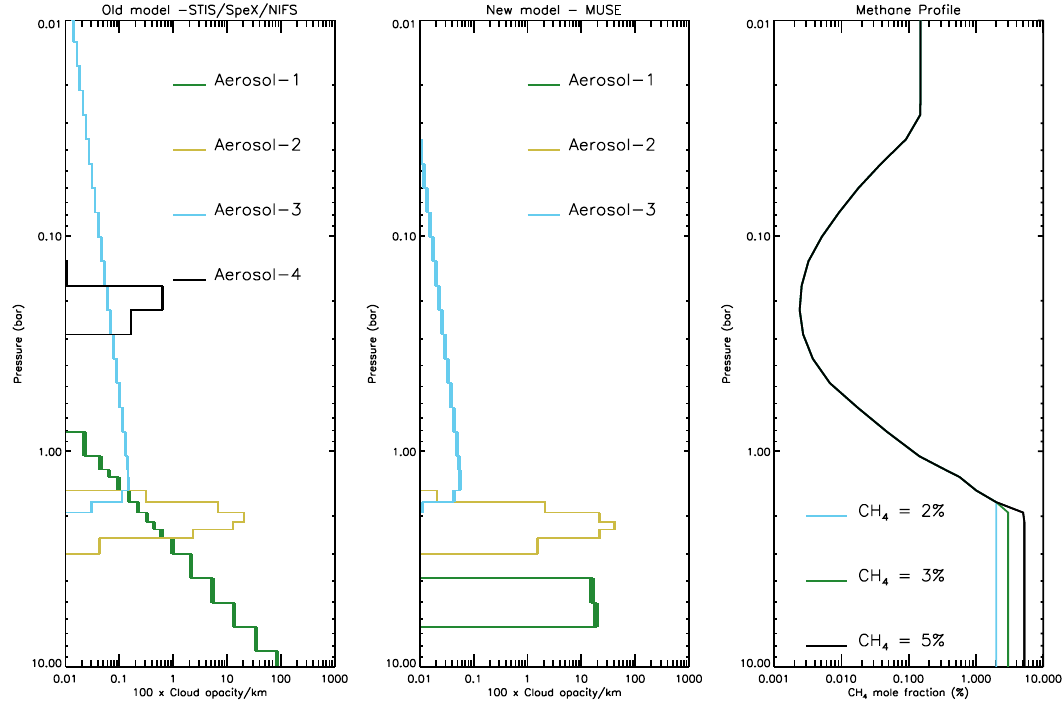}
\caption{Cloud opacity profiles retrieved in this analysis from disc-averaged observations, and assumed example methane abundance profiles.  Left: Cloud scheme used by IRW22 \cite{irwin22} in their combined analysis of HST/STIS, IRTF/SpeX and Gemini/NIFS. Middle: Revised cloud scheme used in this analysis of VLT/MUSE observations. The horizontal thickness of the lines used to plot the profiles indicates the formal retrieved error ranges for the cloud profiles. Right: Assumed methane mole fraction profile for different values of the deep abundance, where the relative humidity above the condensation level is fixed to  50\% and the stratospheric mole fraction is limited to $1.5\times 10^{-3}$ as described in the main text.}\label{fig_cloud}
\end{figure*}

We studied the VLT/MUSE data in the same way as IRW22 and concentrated on the best-resolved, deconvolved `Obs-6' cube, described by \citeA{irwin23}. For this study we used the same simple `step' model for the methane profile as IRW22, with variable deep abundance, a fixed relative humidity above the condensation level (discussed later) and limiting the mole fraction in the stratosphere to $1.5\times 10^{-3}$ \cite{lellouch10}. Sample profiles using this model are shown in Fig. \ref{fig_cloud}. We first analysed the disc-average of the deconvolved `Obs-6' cube, and then analysed the data in latitude bands of width 10$^\circ$, spaced every 5$^\circ$ to achieve Nyquist sampling (masking out discrete clouds in both cases). The high spatial resolution of these observations, combined with a remarkably low level of cloud activity at this time \cite<e.g.,>{chavez23b}, meant that the few high-altitude discrete methane ice clouds that were visible were easily masked-out in our data. This meant that we did not need to include the upper troposphere ($\sim$ 0.2 bar) layer of micron-sized methane ice `Aerosol-4' particles in our model, which only have a significant contribution at wavelengths greater than 1 $\mu$m anyway. Using a Minnaert analysis we extracted the disc-average nadir reflectance spectrum, $(I/F)_0(\lambda)$, and limb-darkening spectrum, $k(\lambda)$, from the observations, where $\lambda$ is wavelength. These spectra were used to reconstruct two synthetic observation spectra, calculated for back-scattering conditions, with the zenith angles (both viewing and solar) set to either 0$^\circ$ (i.e., nadir), or 61.45$^\circ$, which were then fitted simultaneously with our NEMESIS radiative transfer and retrieval tool \cite{irwin08}. We fitted to these two angles simultaneously in order to fit both the mean reflectance and limb-darkening spectra, and these two particular zenith angles were chosen to coincide with two of the zenith angles in our matrix-operator multiple-scattering scheme \cite{plass73}, thus avoiding any interpolation errors. In this analysis \cite<following>{irwin23} we increased the mean radius of the Aerosol-1 particles to 0.1 $\mu$m (standard Gamma distribution of sizes with variance $\sigma = 0.05$), consistent with expectation of fog-like particles, and fixed the mean radius of the Aerosol-2 particles to 0.7 $\mu$m ($\sigma=0.3$), which was found by IRW22 to be the mean size of the Aerosol-2 particles that was most consistent with their observations. The size of the Aerosol-3 particles was left unchanged, with a mean radius 0.05 $\mu$m and variance $\sigma=0.05$. When fitting to the disc-averaged MUSE data we found that the estimated random errors of the reconstructed spectra were smaller than the uncertainties in our radiative transfer model arising from parameterisation choices and the assumed gaseous absorption coefficients of \citeA{kark10}. Hence, following IRW22, the errors on these spectra were set to 1/50 of the maximum disc-averaged nadir reflectance within 100 nm of each point in the spectrum. These errors were reduced by a factor of 2 in the 800 -- 900 nm range to ensure a good fit to the region best able to differentiate between cloud and methane abundance. In addition, the nadir reflectivity error was limited to not exceed 1\% to ensure that NEMESIS fitted well to the wavelengths near the peak of reflection at $\sim$ 500 nm. With these errors (set to be the same for both the 0$^\circ$ and 61.45$^\circ$ spectra) we could fit the spectra to $\chi^2/n \sim$3, giving retrieved parameters with meaningful error values. For the shorter wavelengths we also had to account for a small amount of Raman-scattering (where photons absorbed at shorter wavelengths can be scattered to longer wavelengths in hydrogen-dominated atmospheres) and polarisation effects, which we did following the procedures of \citeA{sromovsky05} and \citeA{sromovsky05pol}, respectively, as outlined by IRW22. To account for the Raman scattering, we added 100 points below the MUSE spectrum in each pixel, covering 373 -- 473 nm, which we reconstructed from a disc-averaged Minnaert analysis of the 2003 HST/STIS observations \cite{kark11}, using the fitted Minnaert $(I/F)_0$ and $k$ spectra to simulate the expected spectrum for the observing geometry of each pixel and scaling this to match the MUSE spectrum where they overlap. In our radiative transfer model we simulated the entire spectrum from 373 to 933 nm, including Raman-scattering. Hence, photons Raman-scattered from wavelengths as short as 373 nm were approximated for in the modelled spectra. However, to avoid the retrieval model trying to fit these synthetic 373 -- 473 nm data, the error on these points was set to 100\%. Finally, for consistency with IRW22 and ease of comparison with their HST/STIS retrievals, the MUSE data were analysed at the same set of wavelengths as for the STIS data. The assumed gaseous absorption coefficients and the assumed vertical profiles of temperature and gaseous abundance were also those used by IRW22 and \citeA{irwin23}.

During the initial retrievals from the latitudinally-averaged data, we found that there was considerable degeneracy between the methane abundance and the opacities of the Aerosol-1 and Aerosol-2 layers, which led to these parameters not varying smoothly with latitude, but showing considerable degeneracy, or `cross-talk'. The reason for this can be understood from Fig. \ref{fig_cloud}, which shows the best-fit cloud model of IRW22 determined from their combined Neptune data set. It can be seen that there is considerable overlap between the Aerosol-1 and Aerosol-2 profiles in the 2--3 bar region. Since the modelled spectra were sensitive to both clouds in this pressure region, the two opacities are able to interfere with each other with, for example, one reducing and the other increasing leading to little change in the overall modelled reflectance. Hence, we revised the parameterisation of the Aerosol-1 layer to be like the Aerosol-2 layer, i.e., a single vertically thin cloud, but centred at a higher, fixed pressure of 5 bar (Fig. \ref{fig_cloud}). This parameterisation was found to fit the MUSE data just as well as the original IRW22 parameterisation, but led to much less overlap between the aerosol distributions and thus much less degeneracy between the Aerosol 1 and 2 opacities and methane abundances in the latitudinal retrievals reported later. This revised Aerosol-1 cloud model was also used by \citeA{irwin23} in their analysis of the NDS-2018 and DBS-2019 features in this data set. We note that our data do not allow us to determine whether there is a clear gap between the Aerosol-1 and Aerosol-2 layers, only that this decoupling is necessary for our retrieval approach to be reliable and stable.

\begin{table}[!h]
\caption{Retrieval Models.\label{tbl-2}}
\begin{tabular}{l l l l}
Parameter & Model 1 & Model 2 & Model 3 \\
\hline
Aerosol-1: & & & \\
$\tau_1$  &  variable  & variable & variable\\
$p_1$     &  5 bar & 5 bar & 5 bar\\
$\Delta p_1$ & 0.1 & 0.1 & 0.1\\
$n_{imag}(\lambda)$ (8 wavelengths) & fixed & variable & variable \\
\hline
Aerosol-2: & & &\\
$\tau_2$  &  variable  & variable & 1.75\\
$p_2$     &  variable & variable & 2.1 \\
$\Delta p_1$ & 0.1 & 0.1 & 0.1\\
$n_{imag}(\lambda)$ (8 wavelengths) & fixed & fixed & fixed\\
\hline
Aerosol-3: & & &\\
$\tau_3$  &  variable  & variable & variable\\
$p_3$     &  1.6 bar & 1.6 bar & 1.6 bar\\
FSH & 2.0 & 2.0 & 2.0\\
$n_{imag}(\lambda)$ (8 wavelengths) & fixed & fixed & fixed\\
\hline
Methane: & & \\
Deep VMR &  variable  & variable & variable\\
RH     &  50\% & 50\% & 50\% \\
\hline
Number of variables $n_x$ & 5 & 13 & 11 \\
\hline
\multicolumn{4}{@{}p{0.7\textwidth}@{}}{\footnotesize N.B., $n_{imag}$ tabulated from 0.3 to 1.0 $\mu$m in steps of 0.1 $\mu$m. $\Delta p$ for Aerosol-1, Aerosol-2 is FWHM of Gaussian function in $\ln p$. Aerosol-3 is extended up from $p_3$ with a set fractional scale height (FSH). Note that where $n_{imag}(\lambda)$ is indicated as `fixed', it means that the spectrum is fixed to that retrieved by the disc-average analysis. } 
\end {tabular}
\end{table}

\begin{figure*}[h]%
\centering
\includegraphics[width=1.0\textwidth]{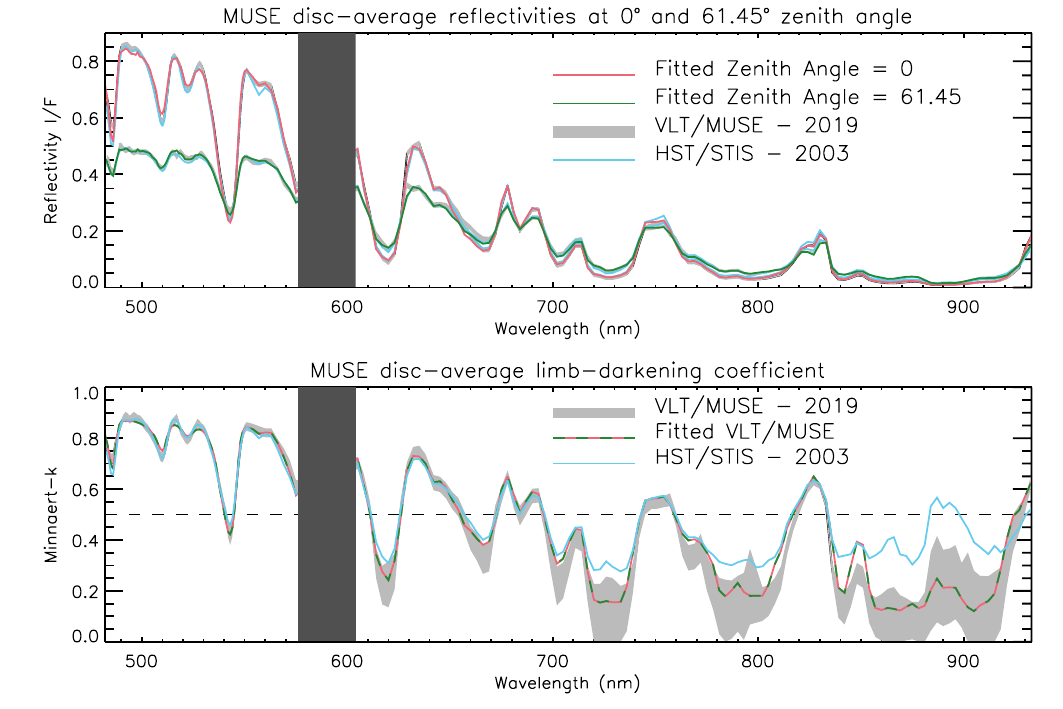}
\caption{Top panel: MUSE spectra reconstructed at 0 and 61.45$^\circ$ zenith angles (viewing and solar) from our disc-average Minnaert analysis, and fits to them using our NEMESIS model. The errors on the synthetic measured spectra are indicated in grey. The spectrum calculated at $\theta=\theta_0=0^\circ$ is of course simply $(I/F)_0$. Bottom panel: MUSE disc-average limb-darkening spectrum $k(\lambda)$ and fitted limb-darkening spectrum, calculated from the NEMESIS fits to the reconstructed spectra at 0 and 61.45$^\circ$ zenith angle using Eq. \ref{eq:recon}. In both panels, the corresponding HST/STIS disc-average spectra from 2003, analysed by IRW22, are over-plotted. The wavelengths greyed-out from 585 -- 602 nm, were reserved for the laser guide star adaptive optics system of MUSE. In the bottom panel, the dashed line at $k$ = 0.5 is added 
to help differentiate between limb-darkened ($k > 0.5$) and limb-brightened ($k < 0.5)$ wavelengths.}\label{fig_muse_spectrum}
\end{figure*}

\begin{figure}[h]%
\centering
\includegraphics[width=0.5\textwidth]{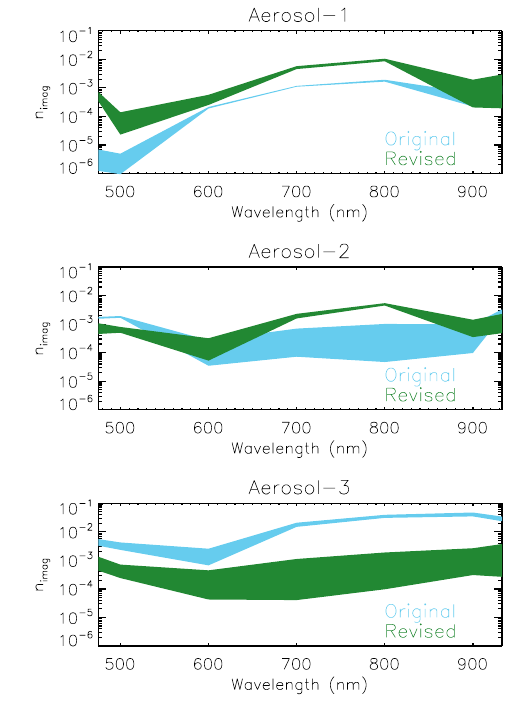}
\caption{Imaginary refractive index ($n_{imag}$) spectra of Aerosol types 1, 2 and 3 derived from fitting the disc-averaged VLT/MUSE observations with our modified aerosol model, compared with the corresponding $n_{imag}$ spectra from the IRW22 analysis of HST/STIS, IRTF/SpeX and Gemini/NIFS Neptune observations \cite{irwin22}.}\label{fig_muse_nimag}
\end{figure}

\section{Results} \label{results}

\subsection{Disc-average retrievals and analysis}

Using the deconvolved `Obs-6' cube, we first determined the disc-averaged limb-darkening properties for each wavelength (masking out discrete cloud features and instrument artefacts) to yield spectra of the fitted parameters $(I/F)_0(\lambda)$ and $k(\lambda)$. These parameters were then used to compute the synthetic measurement spectra seen in Fig. \ref{fig_muse_spectrum}, where the top panel compares two spectra reconstructed from the Minnaert coefficients with both the solar and zenith angles set to either $\theta=0^\circ$ or $\theta=61.45^\circ$, using $(I/F)(\theta) =(I/F)_0 \mu^{2k-1}$ where $\mu_0 = \mu_0 = \cos\theta$. In this plot, the synthetic estimated error limits, discussed earlier, are shaded in grey for both spectra. Note that the spectrum reconstructed at $\theta = 0^\circ$, is simply $(I/F)_0(\lambda)$. Our NEMESIS fits to these spectra, using our modified aerosol model,  are overplotted and show excellent agreement. 

The bottom panel of Fig. \ref{fig_muse_spectrum} compares the limb-darkening coefficient spectrum of the disc-averaged MUSE data with that extracted from our NEMESIS fits to the reconstructed MUSE spectra at $0^\circ$ and $61.45^\circ$, using

\begin{equation}\label{eq:recon}
k_{fit}=0.5 \times \biggl(1 + \frac{\log((I/F)_{fit}(61.45^\circ))-\log((I/F)_{fit}(0^\circ))}{\log(\cos(61.45^\circ))}\biggr),
\end{equation}

showing that NEMESIS correctly fits both the mean reflectivity and limb-darkening spectra. 

The spectra in both panels are compared with those obtained from an identical analysis by IRW22 of the HST/STIS observations of Neptune from 2003 \cite{kark11} and shows that the disc-average data are very similar, which is expected since the MUSE data were scaled to match HST/STIS. However, when comparing the limb-darkening spectra with HST/STIS it can be seen that the MUSE spectra show considerably more limb-brightening at methane-absorbing wavelengths (i.e., $k(\lambda)$ is smaller than for STIS). This may possibly result from the different deconvolution schemes used in the STIS and MUSE datasets, but as we do not have the original undeconvolved STIS observations we are unable to test this. However, this difference in the limb-darkening meant that the Mie-scattering properties calculated from the best-fitting  $n_{imag}$ spectra of the three aerosol layers derived by IRW22 did not achieve a very good fit to these MUSE data. Hence, the $n_{imag}$ spectra had to be refitted to the MUSE observations, which we did using our modified `holistic' aerosol model, also fitting the opacities of the three aerosol layers, the pressure of the Aerosol-2 layer, and the deep methane abundance. The methane relative humidity is difficult to constrain unambiguously from the MUSE wavelength range. IRW22 retrieved a methane relative humidity of 35\% from their data set, which included longer-wavelength IRTF/SpeX observations, while \citeA{kark11} used a fixed value of 60\% to match their HST/STIS data. Here, we fix the methane relative humidity above the condensation level to 50\%.  Our final retrieved disc-averaged $n_{imag}$ spectra for all three aerosols are compared with those derived by IRW22 for Neptune in Fig. \ref{fig_muse_nimag}. Here we can see that for VLT/MUSE the Aerosol-1 particles are determined to be more absorbing (i.e., $n_{imag}$ is higher), the Aerosol-2 particles have similar $n_{imag}$ spectra at short wavelengths, but have higher $n_{imag}$ at $\lambda >$ 600 nm, while the Aerosol-3 particles (most visible at methane-absorbing wavelengths) are required to be considerably more scattering (i.e., $n_{imag}$ is lower) in order to fit the MUSE observations. 

\begin{figure*}[h]%
\centering
\includegraphics[width=1.0\textwidth]{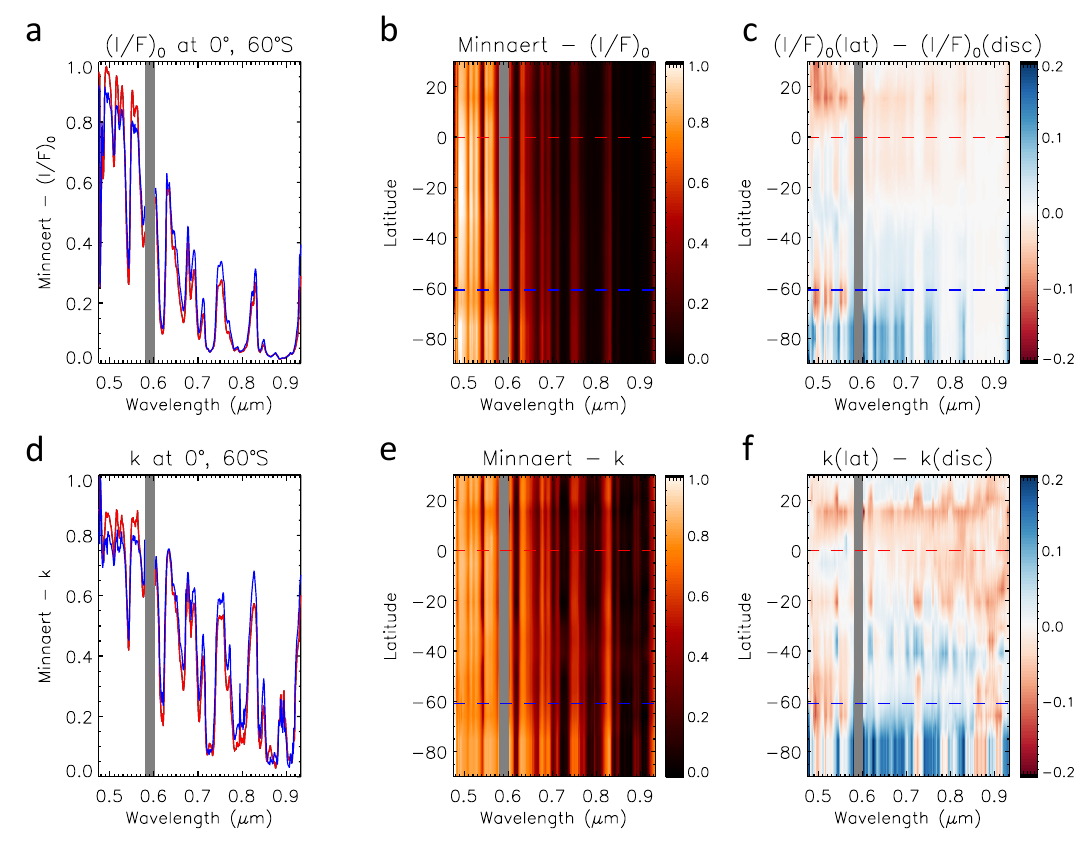}
\caption{Latitudinally-resolved Minnaert limb-darkening analysis of MUSE data. Panel (a) compares the extracted $(I/F)_0$ spectra at the equator (red) and 60$^\circ$S (blue). A contour plot of the extracted $(I/F)_0$ spectra as a function of wavelength and latitude (planetographic) is shown in Panel (b), while Panel (c) shows a contour plot of the difference of $(I/F)_0$ spectra compared with the disc-averaged $(I/F)_0$ spectrum (red regions are darker than disc-average, blue are brighter). In the contour plots the equator and 60$^\circ$S are highlighted with coloured, dashed lines. Panels (d--f) show the same plots as Panels (a--c), but for the fitted Minnaert limb-darkening coefficients, $k$.}\label{fit_minnaert}
\end{figure*}

\subsection{Latitudinally-resolved retrievals and analysis}

Having re-fitted the $n_{imag}$ spectra (and thus all associated Mie-calculated scattering properties) of the three aerosol layers, we then determined latitudinal changes in opacity and methane abundance by analysing spectra reconstructed from the latitudinally-resolved Minnaert analysis of the MUSE dataset, where the data were averaged in latitudes bins of width 10$^\circ$, spaced every 5$^\circ$. A summary of the latitudinally-resolved Minnaert nadir reflectivity spectra, $(I/F)_0(\phi,\lambda)$, and limb-darkening spectra, $k(\phi,\lambda)$, where $\phi$ is the planetographic latitude, is shown in Fig. \ref{fit_minnaert}. For this latitudinal analysis there was not a sufficiently large range of zenith angles south of $70^\circ$S and north of $20^\circ$N to extract meaningful values of both $(I/F)_0(\phi,\lambda)$ and $k(\phi,\lambda)$. Hence for latitudes south of $70^\circ$S, the limb-darkening spectrum, $k(\lambda)$,  was fixed to the average for all latitudes south of 70$^\circ$S and only $(I/F)_0(\phi,\lambda)$ fitted. For latitudes north of $20^\circ$N, in the absence if any other information and since we believe the winds to be, to first order, hemispherically symmetric \cite{sromovsky93}, we assumed north-south symmetry in the cloud distribution also. Hence, the limb-darkening spectra were fixed to those found at the corresponding southern latitude, i.e.,  $k(\phi,\lambda) = k(-\phi,\lambda)$, and again only $(I/F)_0(\phi,\lambda)$ fitted. 
In Fig. \ref{fit_minnaert} we can see the signature of the South Polar Wave (SPW) at $\sim 60^\circ$S in the $(I/F)_0(\lambda)$ difference map at $\lambda <$ 650 nm and also a slight increase in $(I/F)_0(\lambda)$ south of $\sim 25^\circ$S at most wavelengths. Most of the features in the $k(\lambda)$ difference map are in regions of low $k$-values where the reflectances are low and are thus hard to interpret. However, we can see in the SPW at $\sim 60^\circ$S that reflectances are more limb darkened, which is indicative of lower single-scattering albedo. 

With our modified `holistic' atmospheric model parameterisation we then used NEMESIS to fit spectra reconstructed from these latitudinally-resolved Minnaert coefficients at the same two zenith angles used in the disc-average analysis (i.e., 0$^\circ$ and 61.45$^\circ$) and retrieved latitudinal distributions of atmospheric properties using three different model setups. These are summarised in Table \ref{tbl-2} and described in detail in the following subsections.

\begin{figure*}[h]%
\centering
\includegraphics[width=1.0\textwidth]{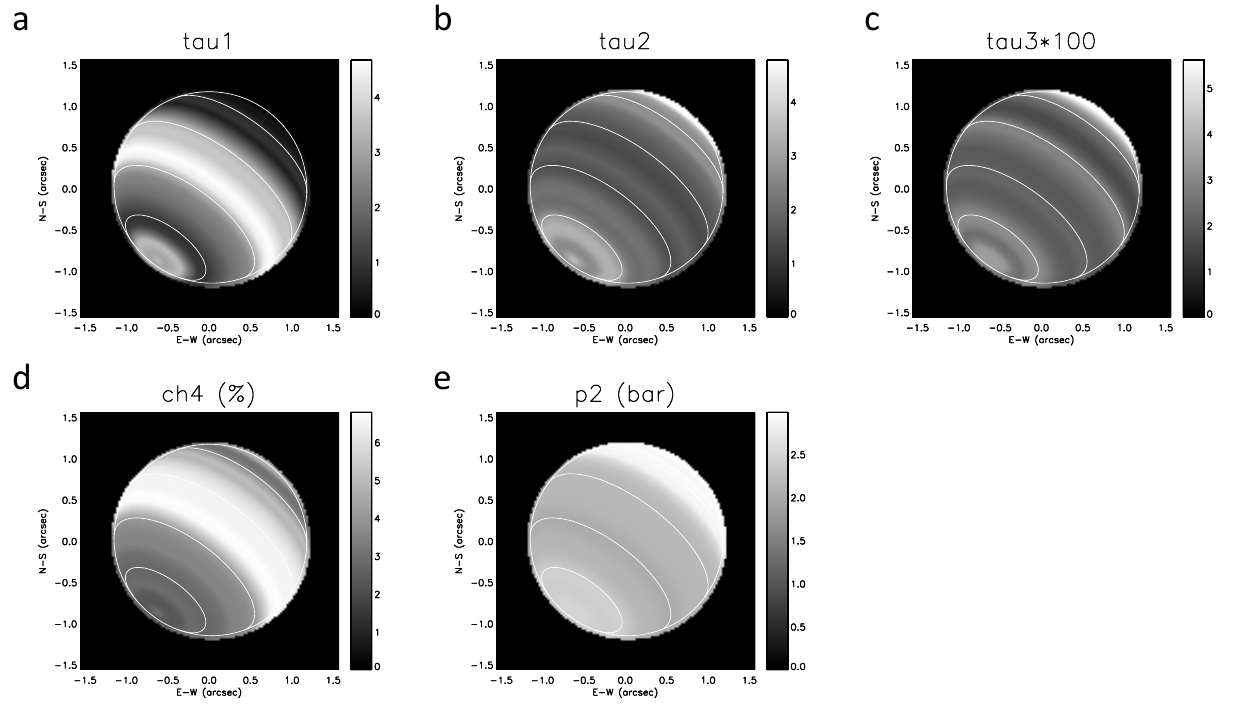}
\caption{Zonally-averaged meridional profiles of atmospheric properties retrieved from the VLT/MUSE Neptune observations (Obs-6) using a model where the particle scattering properties were fixed to the disc-average for all aerosol layers (Model 1). The following fitted properties are projected on to Neptune's disc:  a) Aerosol-1 opacity; b) Aerosol-2 opacity; c) Aerosol-3 opacity ($\times 100$); d) deep methane mole fraction (\%); and e) Pressure of base of Aerosol-2 layer (bar). The cloud opacities referred to here and elsewhere are those at 800 nm. Latitude circles are overplotted for ease of reference, with a spacing of $30^\circ$.}\label{latretrieve1}
\end{figure*}

\begin{figure*}[h]%
\centering
\includegraphics[width=1.0\textwidth]{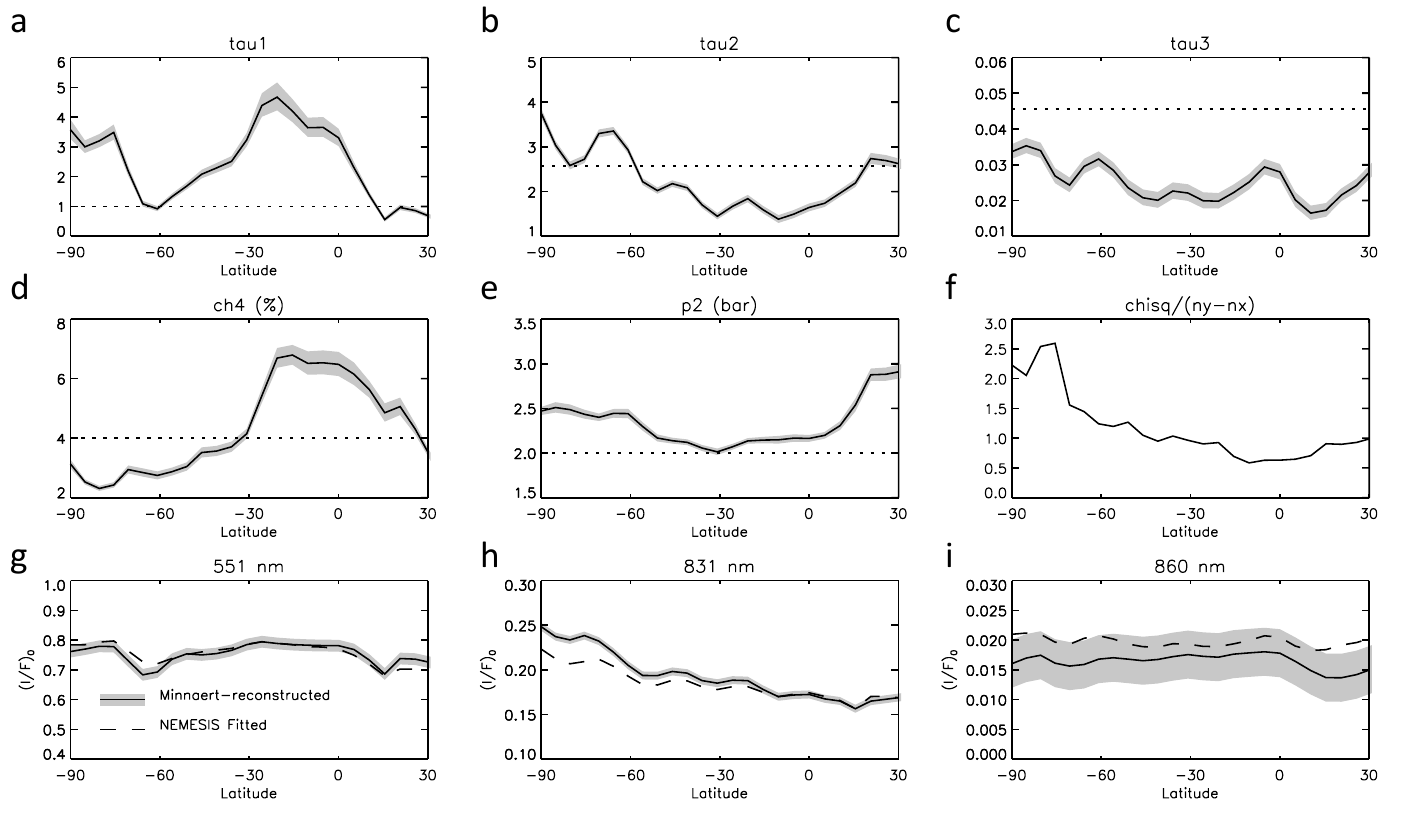}
\caption{Zonally-averaged atmospheric properties retrieved from the VLT/MUSE Neptune observations (Obs-6), using a model (Model 1 of Table \ref{tbl-2}) where the particle scattering properties were fixed to the disc-average for all aerosol layers, showing:  a) Aerosol-1 opacity; b) Aerosol-2 opacity; c) Aerosol-3 opacity ($\times 100$); d) deep methane mole fraction (\%); e) Pressure at the base of Aerosol-2 layer (bar), and f) $\chi^2$/($n_y$--$n_x$) of the fits at different latitudes. The dotted lines in these panels show the \textit{a priori} values assumed. Note that the error bars shown are the formal errors on the retrievals and do not include cross-correlation effects (see Discussion). The bottom row (panels g -- i) shows the latitudinal variations of $(I/F)_0$ at 551, 831 and 860 nm, respectively, where the solid lines show the Minnaert-fitted values (and uncertainties) and the dashed lines show the best-fit values from our retrieval.}\label{latretrieve1plot}
\end{figure*}

\subsubsection{Latitudinally-resolved retrievals with fixed Aerosol-1 $n_{imag}$ spectrum (Model 1).}

Fixing the $n_{imag}$ spectra of the three aerosol types to those determined from the disc-average analysis, we first fitted the latitudinally-resolved reconstructed spectra to retrieve latitudinal distributions of: 1) the opacity of all three aerosol layers (opacities are quoted at 800 nm); 2) the pressure of the Aerosol-2 layer; and 3) the deep methane abundance (Model 1 of Table \ref{tbl-2}), i.e., five parameters in total. The upper tropospheric methane relative humidity was again limited to 50\%.  
In these retrievals the errors on the Minnaert-reconstructed spectra were increased by a factor of 2 to reflect the fact that fewer points were averaged to determine the coefficients and to ensure that the fitted $\chi^2/n$ values were $\sim 1$. The results of this analysis are shown in the form of projected images in Fig. \ref{latretrieve1} and as plots against latitude in Fig.  \ref{latretrieve1plot}. These retrievals show several very clear latitudinal dependencies of the fitted parameters, although the agreement between the observed and fitted reflectivities at 511 and 831 nm is not perfect (Fig. \ref{latretrieve1plot}). The retrieved deep abundance of methane is seen to decrease from 6--7\% at equator to $\sim 3$\% at the pole, a very similar latitudinal dependence to that derived from previous analyses of Neptune's visible/near-IR spectra \cite{kark11,luszcz16, irwin21}, with a boundary at $\sim 20-30^\circ$S. Although the latitudinal dependence of methane abundance is similar to previous estimates, however, the absolute abundances are slightly higher and we note in the Discussion that these depend also on the assumed vertical profile of methane and the modelled aerosol structure, which are here different.
In addition to methane, we also find significant latitudinal variation in the opacity of all three aerosol layers, with variations in the opacity of the Aerosol-2 layer appearing to match the latitudinal variation seen in the 831-nm image (Fig. \ref{museimages}), with opacity peaks at $75^\circ$S, $45^\circ$S and $20^\circ$S.  Higher in the atmosphere, similar, but offset, peaks in opacity can be seen in the retrieved Aerosol-3 opacity, with notable peaks at $80^\circ$S, $60^\circ$S and at the equator. The Aerosol-3 opacity shows some similarity with features in the 848-nm image, which is a wavelength of medium methane absorption, but is better correlated with the features in the 860-nm image, where methane absorption is strong. In particular, the peak in Aerosol-3 opacity at $80^\circ$S corresponds well with a faint ring of brightness near Neptune's south pole at 860 nm and similarly with a brighter zone at the equator. Finally, there is a good correlation between the Aerosol-1 opacity and variations seen at 551 nm, and a sharp reduction in the retrieved opacity near 60$^\circ$S, which coincides with the SPW, with the opacity increasing south of this, corresponding to the brighter reflectivities seen south of the SPW at 511 and 831 nm. Otherwise, the Aerosol-1 opacity and deep methane abundance seem moderately correlated between the equator and 60$^\circ$S, which might suggest upwelling in the 25$^\circ$S -- 25$^\circ$N region, condensing more H$_2$S cloud at $\sim 5$ bar, and downwelling elsewhere. It should also be noted that the spatial variation in the pressure of the Aerosol-2 layer is very small, as expected from previous cloud and methane retrievals for Uranus and Neptune \cite{kark09,kark11}. Finally the $\chi^2/n$ of the fits is good and generally less than $\sim$2, although there are clearly deficiencies at some wavelengths such as 511 and 831 nm. The \textit{a priori} values assumed for the retrieved parameters are overplotted in Fig. \ref{latretrieve1plot} for reference. The \textit{a priori} error for all parameters was set to 100\%, except deep methane abundance, for which a more constrained error of 50\% was assumed.

\subsubsection{Latitudinally-resolved retrievals with variable Aerosol-1 $n_{imag}$ spectrum (Models 2 and 3).} 

\begin{figure}[h]%
\centering
\includegraphics[width=1.0\textwidth]{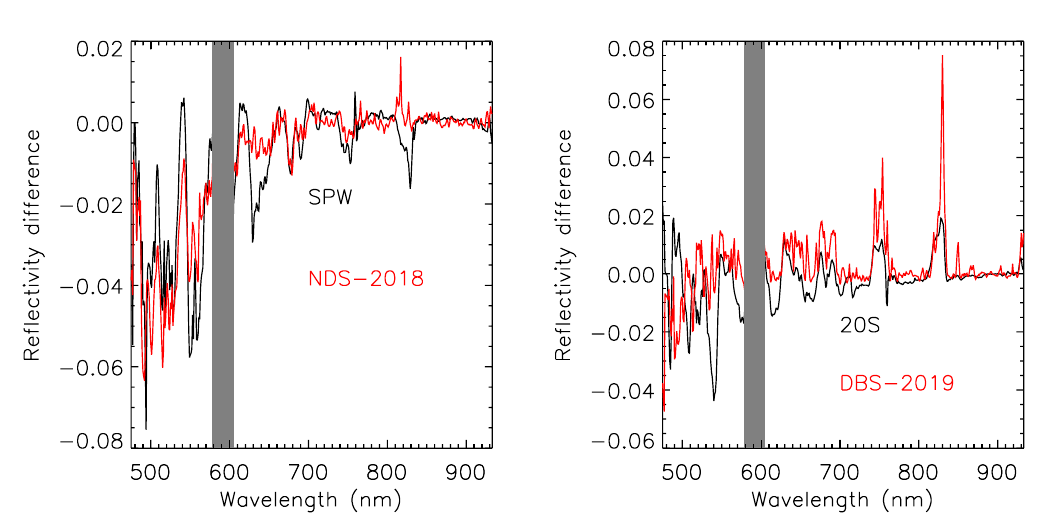}
\caption{Difference spectrum of the South Polar Wave (SPW) at $60^\circ$S (left panel) and the bright deep `zone' at $20^\circ$S (right panel) compared with the difference spectra of the NDS-2018 and DBS-2019 features, respectively, reported by \citeA{irwin23}. The difference spectra of these latitudinal bands is computed as $D(\phi,\lambda) = (I/F)_0(\phi,\lambda) - 0.5\times((I/F)_0(\phi-10^\circ,\lambda) + (I/F)_0(\phi+10^\circ,\lambda))$. The difference spectra of the NDS-2018  and DBS-2019 features are computed as the observed spectra at these locations minus the expected spectra at the same locations calculated from the centre-to-limb reflectivity functions at these latitudes of 15$^\circ$N and 10$^\circ$N, respectively. The comparability between the two sets of difference spectra points to a likely common origin for all these features, namely spectrally-dependent single-scattering albedo spectrum perturbations of the Aerosol-1 layer at $\sim$5 bar.}\label{zone_comparison}
\end{figure}

Although the spectral properties of the SPW at $\sim 60^\circ$S can, to first order, be explained by a thinning of the Aerosol-1 layer, the agreement with the observed reflectivities at 511 nm in Fig. \ref{latretrieve1plot} is not perfect, and the model does not match the 831 nm reflectivities very well. Instead, we wondered if the SPW might be caused by a spectrally-dependent darkening of the Aerosol-1 layer. The values of many properties are varying with latitude towards the pole,
but by subtracting from the $(I/F)_0$ spectrum at $\sim 60^\circ$S the average of the $(I/F)_0$ spectra either side (i.e., those at  $50^\circ$S and $70^\circ$S) we can attempt to isolate the signature of the SPW itself, obtaining the difference spectrum shown in Fig. \ref{zone_comparison}. This spectrum is compared with the difference spectrum of the NDS-2018 dark spot \cite{irwin23}, which is the difference between the observed $I/F$ spectrum in the dark spot and that expected at the same location from the average limb-darkening at this latitude (15$^\circ$N). Figure \ref{zone_comparison} shows that there is considerable similarity between the SPW and NDS-2018 difference spectra, both showing a darkening at wavelengths $<$ 650 nm, with strong methane absorption features indicating the features lie at considerable depth in Neptune's atmosphere. \citeA{irwin23} found that the darkness of the NDS-2018 dark spot could only be explained by a spectrally-dependent darkening of the Aerosol-1 particles, reducing their single-scattering albedo at $\lambda <$ 650 nm, and it is quite possible, and indeed arguably probable, that the SPW darkening is caused by the same mechanism. To test this we repeated our latitudinally-resolved retrievals, but this time we also retrieved the imaginary refractive index spectrum of the Aerosol-1 particles, retrieving this at eight equally-spaced wavelengths from 300 -- 1000 nm, spaced by 100 nm, and with a correlation length of 100 nm (Model 2 of Table \ref{tbl-2}). As noted earlier, the real part of the complex refractive index spectrum was approximately reconstructed with a Kramers-Kronig analysis and the particle scattering properties calculated with Mie theory, smoothing the phase functions with combined Henyey-Greenstein functions.

\begin{figure*}[h]%
\centering
\includegraphics[width=1.0\textwidth]{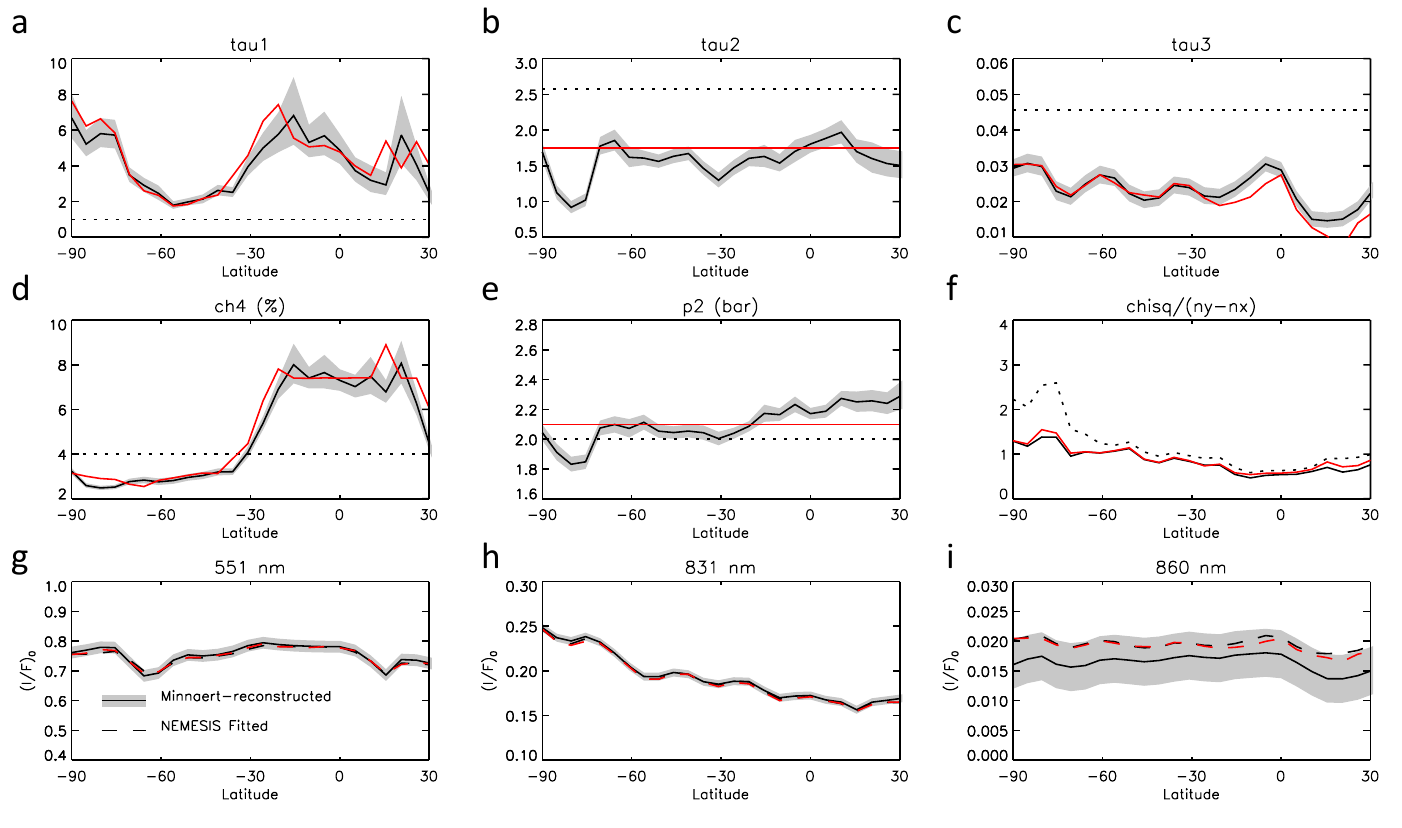}
\caption{As Fig. \ref{latretrieve1plot}, but showing atmospheric properties retrieved from the VLT/MUSE Neptune observations (Obs-6), when the $n_{imag}$ spectrum of the Aerosol-1 particles is also allowed to vary (Model 2). In the middle-right panel, showing the latitudinal (planetographic) variation of $\chi^2/(n_y-n_x)$, the reduced $\chi^2$ for the case where $n_{imag}$ for Aerosol-1 is fixed is overplotted as a dotted line for comparison. Overplotted in red in all panels are the results when the opacity (at 800 nm) of Aerosol-2 is fixed to 1.75 and its pressure fixed to 2.1 bar (Model 3), showing very similar fits to the data and similar reduced $\chi^2$. Note that the error bars shown in panels a -- e are the formal errors on the retrievals  and do not include cross-correlation effects (see Discussion).}\label{latretrieve2plot}
\end{figure*}

\begin{figure*}[h]
\centering
\includegraphics[width=1.0\textwidth]{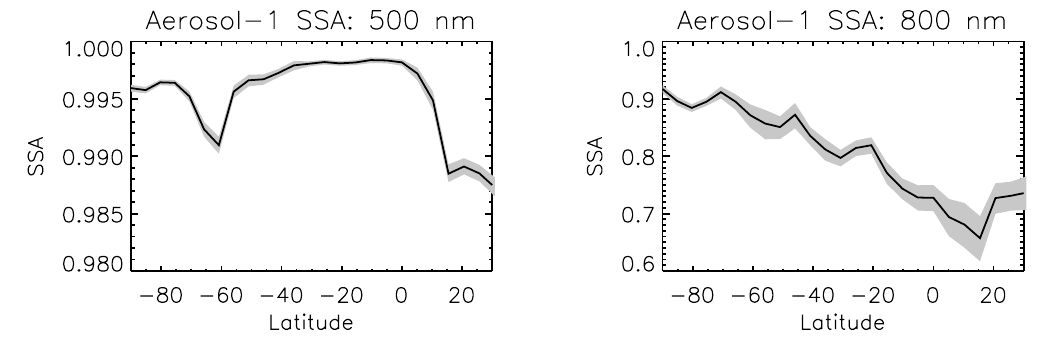}
\caption{Estimated single-scattering albedoes of the best fit Aerosol-1 particles at 500 nm and 800 nm as a function of latitude (planetographic), derived from our fitted $n_{imag}$ spectra using Mie theory. The formal confidence limits are indicated by the grey shading and do not include cross-correlation effects.}\label{ssa_comparison}
\end{figure*}

Allowing the Aerosol-1 $n_{imag}$ spectrum to vary leads to greatly improved fits to the Minnaert-reconstructed reflectivities as shown in Fig. \ref{latretrieve2plot}, with closer fits to the reconstructed nadir reflectances at all wavelengths, including those at 551, 831 and 860 nm, shown here. Although the overall retrievals of Aerosol-3 opacity and methane abundance are similar to those previously determined (Fig. \ref{latretrieve1plot}) the variation in Aerosol-1 opacity shows a less pronounced minimum at $60^\circ$S and the Aerosol-2 opacity and base pressure both seem less correlated with observable variations, although a reduction in Aerosol-2 opacity and pressure level are retrieved at $80^\circ$S. However, the goodness of fit is greatly improved at all latitudes and especially south of $50^\circ$S, with $\chi^2/n$ at $60^\circ$S reducing from 1.23 to 0.99. Of course, the retrieval including $n_{imag}$ has additional model parameters, so we should use the reduced $\chi^2/(n_y-n_x)$ statistic here, where $n_y$ is the number of spectral points and $n_x$ is the number of model parameters. In these retrievals, the total number of fitted spectral points is 374, while the total number of fitted parameters is 5 for the fixed $n_{imag}$ case, and 13 for the variable Aerosol-1 $n_{imag}$ case. This leads to $\chi^2/(n_y-n_x)$=1.24 at $60^\circ$S for the retrieval with fixed $n_{imag}$ and 1.02 for the variable case, which is still clearly a very significant improvement. Fig. \ref{ssa_comparison} shows the latitudinal dependence of the computed single-scattering albedo (SSA) of the Aerosol-1 particles at 500 nm, and we can see a very good correspondence between this and the location of the SPW in the 551-nm image (Fig. \ref{museimages}), with the single-scattering albedo, $\varpi$, of the Aerosol-1 particles significantly reduced at $60^\circ$S. The reduction of $\varpi$ north of $10^\circ$N probably arises due to contamination with the NDS-2018 dark spot not being wholly deconvolved, combined with the fact that the zenith angles are less well sampled at these latitudes and so Minnaert-reconstructed spectra are less reliable. We note here that \citeA{karkoschka11_dark} also inferred a very similar latitudinal variation in the IR single scattering albedo (their Fig. 20).

If the dark SPW at $\sim 60^\circ$S can be explained by a darkening of the Aerosol-1 particles, might a similar process help explain the banded structure seen at 831 nm? The right hand panel of Fig. \ref{ssa_comparison} shows the latitudinal variation of the Aerosol-1 particles' single-scattering albedo at 800 nm. Here, apart from a general trend of increasing $\varpi$ towards the south pole, we see local peaks of $\varpi$ at 0$^\circ$, 20$^\circ$, 45$^\circ$ and 70$^\circ$, which  correspond to peaks in the observed reflectivity at 831 nm (Fig. \ref{museimages}). This suggests that these 831-nm features (and similar banding seen at other longer wavelengths of minimal methane absorption near 680, 750 and 930nm) are caused by a brightening/darkening of the Aerosol-1 particles at these wavelengths. If this is the case, then the effect is very similar to that inferred for the spectral properties of the `Deep Bright Spot', DBS-2019, seen near NDS-2018 and reported by \citeA{irwin23}. To test this hypothesis, Fig. \ref{zone_comparison} compares the difference spectrum at $20^\circ$S (i.e., the difference between the $(I/F)_0$ spectrum at $20^\circ$S and the average of those at $10^\circ$S and $30^\circ$S) with the difference spectrum of DBS-2019. As can be seen there is again a considerable degree of similarity between these difference spectra, with the difference peaks longer than 650 nm being mostly restricted to narrow bands centred on 750 and 830 nm. These spectra are not identical, however, with some differences seen at shorter wavelengths. The DBS-2019 feature is small and it is possible that there remains some spatial mixing of the light from nearby locations left over from incomplete deconvolution. However, the similarity of these spectra at longer wavelengths, especially the narrowness of the reflectivity peaks, leads us to the conclusion that these features likely have a similar cause. Hence, we conclude that the SPW and NDS-2018 features are both caused by a spectrally-dependent darkening of the Aerosol-1 particles at $\lambda < $ 650 nm, while the bright `zones' seen at 831 nm and DBS-2019 are both caused by a spectrally-dependent brightening of the same Aerosol-1 particles at $\lambda > $ 650 nm. Furthermore, the 511 nm and 831 nm features \textbf{must} be caused by spectrally-dependent perturbations of the Aerosol-1 layer, rather than changes in its opacity, since opacity changes would affect both wavelengths while it is clear that the latitude dependence of these features are very different.

Although the formal retrieval errors of the opacity and pressure level of Aerosol-2 shown in Fig. \ref{latretrieve2plot} are smaller than the latitudinal variations, we note that these errors do not include cross-correlation with other model parameters and so may be underestimates. Given that we find that the bright and dark features in our data can mostly be explained by variations in the reflectivity of the Aerosol-1 particles only, we explored whether we could fit the observations by fixing the properties of Aerosol-2 and varying only the other remaining parameters ($\tau_1$, $n_{imag1}(\lambda)$, $\tau_3$, and deep methane abundance). Fixing $\tau_2$ and $p_2$ at all latitudes to the mean values found of 
1.75 (at 800 nm) and 2.1 bar, respectively (Model 3 of Table \ref{tbl-2}), we re-ran our retrieval model, whose results are overplotted in red in Fig. \ref{latretrieve2plot}. The quality of the fits can be seen to be only very slightly reduced, with the same trends seen in all the other parameters.
Hence, we conclude that to first order the Aerosol-2 layer is latitudinally invariant and that the vast majority of the latitudinal changes seen in the MUSE data can be attributed to changes in: 1) the methane abundance; 2) the opacity and scattering properties of the Aerosol-1 layer at $\sim$5 bar; and 3) the opacity of the upper tropospheric Aerosol-3 layer. We will return to discuss retrieval errors in the discussion section.

\section{Discussion}\label{discussion}

The presence of bright `zones' and dark `belts' in the Aerosol-1 layer at wavelengths of minimal methane absorption near 830 nm (and also less clearly at 670, 750, and 930 nm), with the zones separated by $\sim 25^\circ$ is not something that has specifically been noted before and is very curious given that the measured wind speeds show a much more slowly varying latitudinal dependence. These finer-scale latitudinal variations in reflectivity are, in fact, just visible in HST/WFC3 F845M images \cite<e.g.,>{chavez23b}, although much less clear, and at longer wavelengths \citeA{sromovsky14} note finer-scale structure in H-band (1.5 $\mu$m)
Keck observations of Uranus. Although just visible in broader wavelength filters, the banding seen here is only very clear at wavelengths of extremely weak methane absorption and is not seen at wavelengths of stronger methane absorption. Only high-resolution IFU spectrometers such as MUSE can discriminate between these deep features and shallower cloud structures that would appear identical in broader filters, and thus isolate their surprising origin. Such very different structure below the visible haze layers is reminiscent of Cassini/VIMS observations of the 5-$\mu$m emission from Saturn's atmosphere, which revealed finely detailed bright and dark bands that bore little resemblance with the broader structures seen at lower pressures \cite<e.g.,>{baines06}. This suggests that the circulation of Neptune's atmosphere below the 2--3-bar Aerosol-2 ice/haze layer is likely very different from that seen above that layer. 

Analysing these VLT/MUSE data we have shown that Neptune's SPW and dark spots are likely to be caused by the same process, i.e., a darkening of the particles in the  $\sim$5-bar-Aerosol-1 layer at wavelengths less than 650 nm. Our analysis of the SPW signature is helped by the zenith-angle coverage allowing us to constrain the limb-darkening properties, but is complicated by the difficulty in extracting the SPW signature from the  latitudinal variations in other parameters. In contrast, the NDS-2018 difference spectrum is relative to other locations in the same latitude band, so extracting its signature is not complicated by latitudinal variations. However, during the time that NDS-2018 was visible in our data set, from `Obs-6' and `Obs-10', the observed zenith angle only increased from 42.36$^\circ$ to 45.37$^\circ$ and thus we do not have sufficient zenith angle data to constrain its limb-darkening characteristics. However, IRW22 analysed the limb-darkening properties of both the SPW and GDS from Voyager-2 imaging observations of Neptune in 1989 at several wavelengths and found they had similar limb-darkening, which again points to a likely common cause of these short-wavelength dark features. At longer wavelengths, the signature of the deep bright spot DBS-2019 is similar to the differences between the bright `zones' and dark `belts' seen at longer continuum wavelengths, such as 831 nm, and both seem to be caused by a brightening of the particles in the $\sim$5 bar-Aerosol-1 layer at wavelengths $> 650$ nm. It is intriguing that two very distinctive and very different features are caused by modifications of the same Aerosol-1 layer at wavelengths either less than or greater than 650 nm, and we thus looked to determine whether there might be a simple explanation for these very different reflectance signatures.

\begin{figure}[h]
\centering
\includegraphics[width=0.6\textwidth]{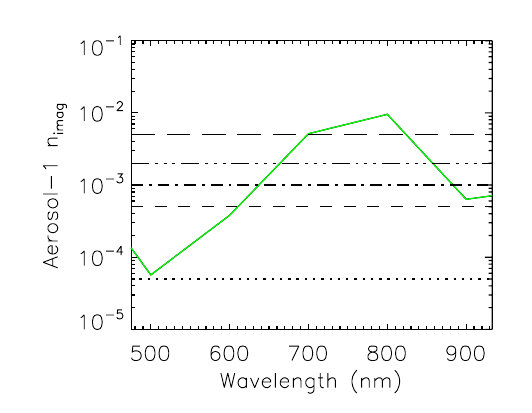}
\caption{Imaginary refractive index spectrum of the fitted disc-averaged Aerosol-1 particles (green), compared with the $n_{imag}$ spectra of the additional particles added to the Aerosol-1 layer in the reflectivity calculations shown in Fig. \ref{refl_diff}.}\label{nimag_plot}
\end{figure}

In Fig. \ref{nimag_plot} we have again plotted the $n_{imag}$ spectrum of the Aerosol-1 particles retrieved from our disc-average Minnaert limb-darkening analysis. As can be seen we retrieve low values near 500 nm, making the particles highly scattering ($\varpi = 0.999$), rising to large values at 800 nm, making the particles poorly scattering ($\varpi = 0.777$). The increase of $n_{imag}$ with wavelength is commonly seen with photochemically-produced particles as discussed by IRW22, who conclude that the particles in Aerosol-1 layer are likely composed of a mixture of bright H$_2$S ice particles and photochemical products mixed down from above. The large single-scattering albedo difference on either side of 650 nm is reminiscent of the difference between deep dark and bright spots and we wondered what would happen to the computed disc-averaged $(I/F)_0$ spectrum if we added an additional component of particles to the Aerosol-1 layer with different, fixed $n_{imag}$ spectra. The $n_{imag}$ values we chose were $5\times 10^{-5}$, $5\times 10^{-4}$, $5\times 10^{-3}$, $2\times 10^{-3}$ and $1\times 10^{-3}$ (overplotted in Fig. \ref{nimag_plot}), roughly spanning the range of $n_{imag}$ seen in the retrieved Aerosol-1 spectrum. We also tested whether the particle size might be important and used a size distribution for the additional particles either equal to that assumed for the Aerosol-1 layer with mean radius $r_0$ = 0.1 $\mu$m, or increased to $r_0$ = 1.0 $\mu$m. The particle size variance in both cases was set to $\sigma=0.05$. The additional particles had the same vertical distribution as the existing Aerosol-1 layer and had an opacity (at 800 nm) relative to the existing Aerosol-1 layer of 10\% for the 0.1-$\mu$m particles or 30\% for the less back-scattering 1.0-$\mu$m particles. 

\begin{figure}[h]
\centering
\includegraphics[width=0.5\textwidth]{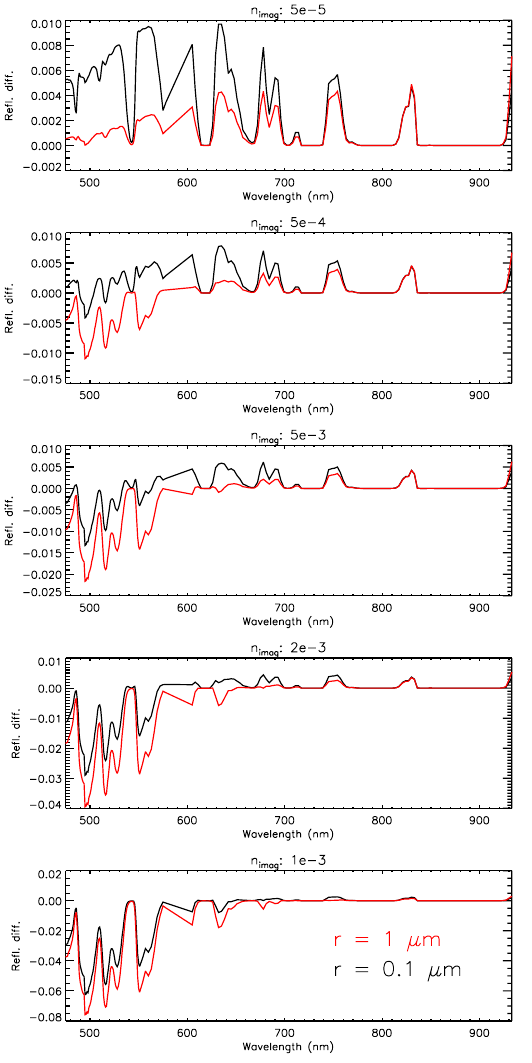}
\caption{Change in calculated disc-averaged $(I/F)_0$ spectrum when an additional opacity of particles with specified $n_{imag}$ is added to the particles in the Aerosol-1 layer. The additional particles either have a mean radius of 0.1 $\mu$m (black) or 1.0 $\mu$m (red) and the additional opacity (at 800 nm) is 10\% or 30\% of the existing Aerosol-1 opacity, respectively. In both cases the variance of the additional size distribution in 0.05.}\label{refl_diff}
\end{figure}

The differences in the computed $(I/F)_0$ spectra are shown in Fig. \ref{refl_diff}, which demonstrates that when the additional particles have larger $n_{imag}$ values, and thus lower single-scattering albedo, we see a darkening at shorter wavelengths, but little effect at longer wavelengths.  This is understandable given that the additional darker particles will have a large effect at wavelengths where the surrounding Aerosol-1 particles are highly scattering, but will have a minimal impact at wavelengths where the surrounding particles are already rather absorbing. Similarly, adding highly-scattering particles to the Aerosol-1 layer, with low $n_{imag}$, will not make much difference at wavelengths where the particles are already highly scattering, but will significantly affect wavelengths where the surrounding particles have lower albedo. For either additional particle size, when $n_{imag}$ is high (e.g., $1\times 10^{-3}$), the difference in the computed $(I/F)_0$ spectrum matches closely to the SPW and NDS-2018 difference spectra (Fig. \ref{zone_comparison}). On the other hand, for highly scattering particles (e.g., $n_{imag} = 5\times 10^{-5}$) we see not only narrow reflectance peaks at longer wavelengths, but also an increase in reflectance at shorter wavelengths. However, for small particles and $n_{imag}$ in the range of $5\times 10^{-4}$ to $5\times 10^{-3}$ the effect on the $(I/F)_0$ spectra is very similar to the $20^\circ$S and DBS-2019 difference spectra of Fig. \ref{zone_comparison}, with narrow reflectance peaks at longer wavelengths and little clear signature at shorter wavelengths. Hence, rather than trying to explain why the average scattering spectra of the Aerosol-1 particles in the 831-nm bright regions and 500-nm dark spots are so different from the background, we could instead consider simpler solutions where dark spots are caused by the addition of dark particles, or `chromophores', that are generally absorbing across the MUSE range and 831-nm bright regions are caused by the addition of other particles that are generally scattering, such as fresh H$_2$S ice. 

Although this chromophore model is simple and easy to understand, it does then leave the question of what the additional chromophores might be. An alternative explanation for dark regions in Neptune's atmosphere, noted by IRW22, is that warm regions at the Aerosol-1 pressure level might lead to H$_2$S ice sublimating off from the particles, revealing their darker photochemically-produced cores. This would have more of an effect at wavelengths where the reference Aerosol-1 particles are bright. Equally, cooler regions might result in more  H$_2$S ice condensing on to the particles, which would not make them more reflective at wavelengths where they are already bright, but could lead to them becoming brighter at the longer wavelengths where we see belts and zones. Observations of Neptune with VLA and ALMA \cite{tollefson21} note latitudinal variations in brightness temperature at wavelengths sounding 4 -- 8 bar that are similar to the reflectivity variations seen in our longer wavelength ($>$ 600 nm) methane windows, which suggests there might indeed be a link. However, the direction of such a link is arguable: warmer regions might lead to H$_2$S ice sublimating off the haze cores to make the particles darker, but it might also be that darker aerosol particles will absorb more sunlight and thus increase the local atmospheric warming.

\begin{figure*}[h]%
\centering
\includegraphics[width=1.0\textwidth]{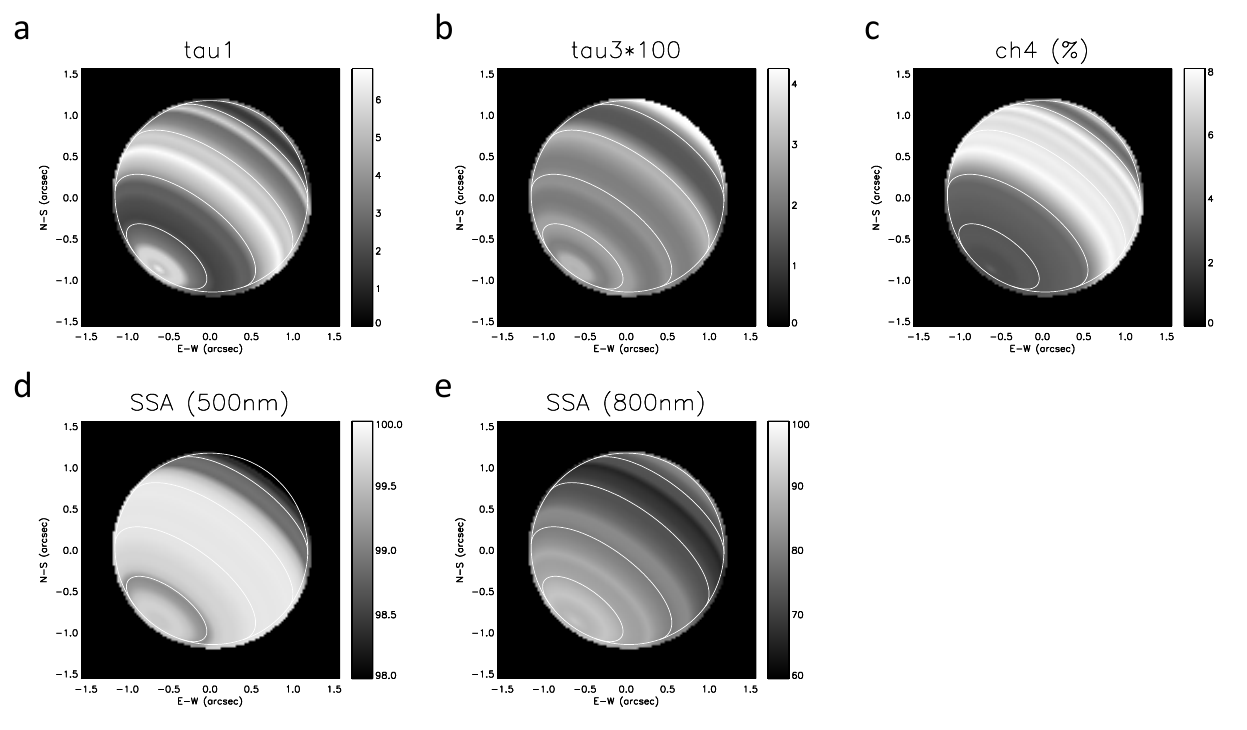}
\caption{Similar to Fig. \ref{latretrieve1}, but showing atmospheric properties retrieved and projected on to Neptune's disc when the Aerosol-1 $n_{imag}$ spectrum is allowed to vary, and the Aerosol-2 properties are fixed (Model 3), mapping:  a) Aerosol-1 opacity; b) Aerosol-3 opacity ($\times 100$); c) deep methane mole fraction (\%); d) Aerosol-1 single-scattering albedo (\%) at 500 nm; and e) Aerosol-1 single-scattering albedo (\%) at 800 nm. The cloud opacities are those at calculated at 800 nm. Latitude circles are overplotted for ease of reference, with a spacing of $30^\circ$. 
}\label{latretrieve2}
\end{figure*}

While the `zones' at 831 nm are consistent with a brightening of the Aerosol-1 particles, we can see from Fig. \ref{ssa_comparison} that there is general increase of the 800-nm single-scattering albedo (SSA) towards the south pole. In Fig. \ref{latretrieve2} selected retrieved parameters (for Model 3, where the Aerosol-1 $n_{imag}$ spectrum was allowed to vary, but the opacity and pressure level of the Aerosol-2 layer fixed) are plotted as disc images. The SSA image at 500 nm shows generally good correspondence with the brightness of the 511-nm image in Fig. \ref{museimages} and the  SSA image at 800 nm shows good correspondence with the 831-nm image south of the equator, except north of the equator, where the Minnaert-reconstructed spectra are less reliable. In particular, the bright 831-nm zone seen at $75^\circ$S (which was noted by \citeA{irwin23}), just south of the SPW corresponds with a region of high 800-nm SSA, and another such region lies on the SPW's northern edge. Might it be that fresh material is upwelling on either side of the SPW, then moves into the SPW, suffering photochemical alteration along the way and darkening it to form the 500-nm-dark SPW feature? A possible candidate for the fresh material might be H$_2$S ice, and gaseous H$_2$S was detected in Gemini/NIFS observations of Neptune in 2010 \cite{irwin19h2s}, who found its signature (near 1.5 $\mu$m) to be stronger towards Neptune's south pole, indicating higher abundance at 2--3 bar. However, deeper in the atmosphere, observations with the Very Large Array and ALMA \cite{tollefson21} find low thermal emission at equatorial latitudes and high thermal emission at polar latitudes. At the depths sounded at these microwave wavelengths ($\sim$10 bar), low thermal emission regions are interpreted as being regions of high H$_2$S abundance and thus \citeA{tollefson21} concluded that H$_2$S  is enriched at equatorial latitudes, just as we see in our CH$_4$ retrievals, indicative of strong and persistent upwelling. It may be that the H$_2$S signature seen by Gemini/NIFS was obscured near the equator by the enhanced Aerosol-1 opacity we retrieve in this study, making it easier to detect at more polar latitudes near 1.5 $\mu$m. Or is it the case that the latitudinal distribution of H$_2$S is different at different pressure levels?

Much higher in the atmosphere the ring of stratospheric haze (Aerosol-3) surrounding the south pole at 80$^\circ$S is intriguing.  We can see the feature in our MUSE data at all methane-absorbing wavelengths longer than 650 nm and also searched for it in HST/WFC observations made during the same apparition, detecting it in the FQ619N and FQ727N filters (both narrow-band filters, centred on wavelengths of strong methane absorption), but not finding it in the broadband HST/WFC3 filters.  The ring of material clearly resides high in the atmosphere, near the tropopause and perhaps marks the edge of the south polar vortex in Neptune's upper atmospheric  circulation, although \citeA{fletcher14} and \citeA{depater14} put the edge of the south polar vortex, or the prograde polar  jet, at $\sim$ 70$^\circ$S. In addition to the bright ring at 80$^\circ$S at methane-absorbing wavelengths we also see a slight increase of reflectivity at the equator. Such a distribution indicates upwelling of air at mid-latitudes followed by photodissociation of methane leading to haze production, with the air then moving the haze north and south and concentrating the aerosols at the equator and poles before it downwells \cite{depater14,tollefson21}.

The retrieved `deep' abundance of methane varies from 6--7\% at the equator to $\sim$3\% at polar latitudes, with a boundary at 20--30$^\circ$S. These abundances are slightly higher than previously retrieved \cite{kark10,irwin19methane,irwin21}, but we know that the absolute abundance depends both on the assumed vertical profile of methane and also the assumed vertical profile of clouds/hazes. Our scheme has a methane profile that is fixed up to the condensation level, falling with a prescribed relative humidity above,  and is combined with three simply vertically parameterised aerosol layers. Although we show that this model matches the data very well, alternative parameterisations could fit similarly well, but return different absolute methane abundances. In particular, \citeA{kark09, kark11} and \citeA{sromovsky11} favour a `descended' methane profile that is depleted in the lower trosposphere (at 1-5 bar), but returns to the same value at all latitudes at great depth. Indeed, if latitudinal variations in methane really extended to great depth, then \citeA{sromovsky11} notes that there would be a significant gradient of mean molecular weight with latitude and significant gradients of vertical wind shear \cite<e.g.,>{sun91,tollefson18} that are inconsistent with the observed winds. Hence, while we are confident of the shape of the retrieved latitudinal abundance of methane, we can be less sure of the absolute profile and abundances. However, the general increase at longer MUSE wavelengths of the latitudinally-resolved Minnaert nadir reflectivity $(I/F)_0$ south of $20-30^\circ$S, relative to the disc-averaged $(I/F)_0$ spectrum, noted in Fig. \ref{fit_minnaert}, can now simply be interpreted as being caused by the lower methane abundances determined at these latitudes. 

The circulation of Neptune's atmosphere has been studied for decades, but is still puzzling. In the horizontal direction, we know that the winds are predominantly zonal (i.e., east-west), with strongly retrograde winds at the equator with speeds of nearly 400 m/s, becoming prograde polewards of $\sim$45$^\circ$N and $\sim$45$^\circ$S, reaching peak prograde speeds of over 200 m/s at $\sim$75$^\circ$N and $\sim$75$^\circ$S, before falling to zero at the poles \cite{sromovsky93, sanchez19}. In the vertical direction, the presence of thick upper-tropospheric methane ice clouds and cooler retrieved upper-tropospheric temperatures at mid-latitudes (20--40$^\circ$N and 20--40$^\circ$S)  indicates upwelling at these latitudes and cooling as the air adiabatically expands at the tropopause and moves towards the equator and poles \cite{conrath91,bezard91,fletcher14,depater14,roman22}. This circulation is consistent with the distribution of Aerosol-3 particles we detect in our MUSE observations. Deeper down, the significant vertical reduction of the mean molecular weight at the CH$_4$ condensation level at $\sim$ 2 bar seems to act as a barrier to vertical motion (e.g., IRW22) and below this the vertical circulation appears to switch with air rising at equatorial latitudes and falling nearer the pole, leading to higher equatorial abundances of CH$_4$, seen here, and also high deep H$_2$S abundances seen at microwave wavelengths \cite{tollefson21}. This simple `stacked circulation' view \cite<e.g.,>{tollefson19, fletcher20}, however, is inconsistent with the banded structure seen here at methane continuum wavelengths longer than 650 nm and may also be inconsistent with retrieved H$_2$S abundances (at 1.5 $\mu$m) at pressures less than 5 bar \cite{irwin19h2s}, suggesting that the circulation may actually be even more complicated than previously thought. An interesting analogy is the distribution of NH$_3$ seen in Jupiter's deep atmosphere by the Juno spacecraft \cite{li17}, which appears to rise in an equatorial plume, with much greater abundances than at mid-latitudes, although a more zonal NH$_3$ structure is seen at the 0.5--2-bar level by both Juno \cite{fletcher21}, and also in mid-infrared observations \cite<e.g.,>{fletcher16}; this is believed to arise and be maintained by a similar double `stacked cell' system \cite<e.g.,>{ingersoll00, showman05, duer21,fletcher21,moeckel23,depater23}.

The data analysed in this study have been fitted by varying just the following few variables: the opacity of the three aerosol layers, the pressure level of Aerosol-2 layer, the `deep' methane abundance, and the imaginary refractive index spectra of the three aerosol types. The scattering properties of a distribution of particles in a planetary atmosphere depend on very many factors, including the size distribution, composition, and  mix of different particle species. Regrettably, for Neptune we have very little \textit{a priori} constraint on what the expected particles and their size distributions should be, nor any reliable information on their complex refractive index spectra. As a result, have to try and retrieve these from the data themselves if we are fit the observations satisfactorily. We do this using optimal estimation method \cite{rodgers00}, expanded upon below, which is a deepest descent approach and thus is most stable when the rate of change of radiance with model parameter is a smoothly varying function of the value of that parameter. Unfortunately, we find that the rate of change of radiance does not vary smoothly enough with the mean radius of the size distribution to reliably retrieve this with our model. Hence, instead we conduct separate retrievals for a range of fixed particle size distributions and choose those that match the data best (e.g., IRW22). We do find, however, that the model behaves well when varying the $n_{imag}$ spectra, and we do find that we need for the scattering properties of the particles to vary more rapidly with wavelength than simple changes in $r_{mean}$ or the variance of the size distribution can achieve. We attribute this to the fact that the particles are likely made up of unknown photochemical products that have strong absorption bands. We thus fix the size distribution of the particles in a certain aerosol layer to an acceptable shape (compared with a range of observations) and then retrieve the $n_{imag}$ spectra, reconstructing the $n_{real}$ spectra using the Kramers-Kronig approach. In reality the particles may be a mix of separate ice and photochemical particles, but they could also be combined together by ‘riming’ or some other combination mechanism. Unfortunately, we just do not have this information and so we instead retrieve the mean scattering properties of the layer by fitting $n_{imag}$ as we describe. Although an approximation, we find that our approach allows us to fit the data well and we can then work on interpreting what we have found. This is precisely the point of the first part of this discussion, where we show that the variation of the $n_{imag}$ spectra with latitude of the ‘mean’ particles in the Aerosol-1 layer  can be interpreted as being due by the presence of a background distribution of particles, presumably rich in photochemical haze, combined with a latitudinally-varying opacity of fresh ice crystals. It may be that these two particle distributions have different mean radii and variances, but these cannot be uniquely separated using these data and so we assume a single mean size distribution. Ideally, we would have the complex refractive index spectra of a set of laboratory-measured possible condensates to test against the measured spectra, but this data set does not yet exist. In the meantime, we believe that we have developed a method that reliably retrieves the mean scattering properties of the aerosol layers, which can then be interpreted to gain novel insights into the hazes and clouds in Neptune’s atmosphere. 

Finally, we note that the retrievals presented here were conducted using the optimal estimation \cite{rodgers00} framework of the NEMESIS model \cite{irwin08}. To model these spectral observations requires a full multiple-scattering radiative transfer model, which must be run at many latitudes, many wavelengths, and with many tunable parameters. Hence, this model is computationally very expensive and to combine it with a Bayesian retrieval framework (such as Monte-Carlo Markov Chain or Nested Sampling, e.g., \citeA{skilling06}), which require tens of thousands of iterations, would be prohibitively computationally expensive. However, although cross-correlation between parameters can to be interpreted in optimal estimation method from the computed covariance matrix, this is much less easy to comprehend and present than the  `corner plots' of Bayesian approaches. The retrieval errors we present in this work are derived from diagonal elements of the retrieved covariance matrix and corrected for the original \textit{a priori} errors, since if the retrieved error is the same as the \textit{a priori} error we have not learnt anything new. Hence, the errors shown are $\sigma=1/\sqrt(1/\sigma_\mathrm{ret}^2-1/\sigma_\mathrm{apr}^2)$, where $\sigma_\mathrm{ret}^2$ are diagonal components of the retrieved covariance matrix and $\sigma_\mathrm{apr}^2$ are diagonal components of the \textit{a priori} covariance matrix. However, these errors still do not wholly account for cross-correlation effects. Instead, we estimate the magnitude of these effects by fixing some parameters and observing the effect on the other retrieved variables. For example, going from Model 2 to Model 3 we show that by fixing the pressure and opacity of the Aerosol-2 layer we can achieve very similar quality fits and very similar variations with latitude of the other retrieved parameters. How, then, can we be sure that the retrieved latitudinal variations of the remaining parameters in Model 3 are reliable? 
Figure \ref{zone_comparison} shows that dark spots, the SPW and the zone/belt signature at longer continuum wavelengths are all most consistent with changes in the Aerosol-1 layer, and that this study and previous analyses \cite{irwin22,irwin23} find that spectrally-dependent changes in the reflectivity of the Aerosol-1 particles provide the best solution, which is what Model 3 achieves (Fig. \ref{ssa_comparison}). There are no clear latitudinal variations that have a spectral signature consistent with significant latitude variations in the Aerosol-2 layer, but variations seen at methane-absorbing wavelengths, and which thus must be high in the atmosphere, are matched by latitude variations in the opacity of the Aerosol-3 layer and there is a clear methane variation signature in the observations, noted earlier. Hence, our Model 3 is consistent with all the evidence in the MUSE data set and is, we believe, reliable. We also note that as this model was based on the `holistic' model of IRW22, which was originally derived from a much wider wavelength range data set, it is likely to be applicable to longer wavelengths also. 
Observations of Neptune have recently been made  with the NIRSpec and MIRI instruments on the James Webb Space Telescope (JWST) and we look forward to seeing if our parameterisation can be extended successfully to the longer wavelengths sampled by JWST also.

\section{Conclusions}\label{conclude}

Our analysis of latitudinally-resolved centre-to-limb spectra of Neptune, observed with VLT/MUSE in October 2019, has revealed new constraints on the mechanism that causes the darkening of the South Polar Wave at $\sim$ 60$^\circ$S, which we conclude is common with the darkening mechanism for discrete dark spots such as Voyager-2's Great Dark Spot and the more recent NDS-2018 feature. Using a modified version of the `holistic' aerosol model of IRW22 we find that both these features are consistent with being caused by the darkening (at wavelengths less than 650 nm) of particles in an Aerosol-1 layer, based at $\sim$5 bar, which we suggest is composed of a mixture of photochemically-produced haze, mixed down from their production level above, and H$_2$S ice. In addition, we note a latitudinal brightening and darkening (with a scale of $\sim$ 25$^\circ$) in Neptune's aerosols, which is only visible at regions of very low methane absorption longer than 650 nm. We show that this feature is caused by a brightening (at wavelengths greater than 650 nm) of the particles in the same Aerosol-1 layer at $\sim$ 5 bar. We conclude that these different features seen at $\lambda <$ 650 nm and $\lambda >$ 650 nm must be caused by spectrally-dependent perturbations of the Aerosol-1 scattering properties, rather than opacity changes of this layer, since they have very different spatial structures and changes in the opacity would affect both wavelength ranges similarly. 

Above the Aerosol-1 layer at $\sim$5 bar, we find the properties of the main Aerosol-2 layer, confined to a region of high static stability at the CH$_4$-condensation level at $\sim$2 bar, is to first order invariant with latitude, although the base pressure and opacity may be slightly reduced at 80$^\circ$S. Even higher in the atmosphere, variations in the opacity of the Aerosol-3 tropospheric haze are able to reproduce the observed variation in reflectivity at methane-absorbing wavelengths. We find the abundance of this aerosol to be higher at the equator and also in a narrow `zone' at $80^\circ$S, which supports the hypothesis that at the tropopause air is rising at mid-latitudes and then moving polewards and equatorwards, concentrating the photochemical haze products there. The bright ring at 80$^\circ$S would appear to define the edge of a polar cyclone that has been observed on Neptune for the past two decades, and which are often seen in giant planet atmospheres.

The detailed conclusions of this paper are:

\begin{enumerate}
    \item We find a slightly modified version of the `holistic' aerosol model of IRW22, consisting of three distinct layers, fits the VLT/MUSE data very well, with the Aerosol-1 parameterisation revised to be a single, vertically-confined layer;
    
    \item We find similar, but revised $n_{imag}$ wavelength dependences to those derived by IRW22 for the three aerosol types. This revision was necessary to match centre-to-limb functions of the deconvolved MUSE observations. In particular, we see more limb-brightening at methane-absorbing wavelengths in our deconvolved MUSE data, requiring a higher single-scattering albedo for the upper atmospheric haze particles in the Aerosol-3 layer;
    
    \item We find that the Aerosol-1 opacity varies strongly with latitude, with high values retrieved at the equator and south pole and lower values retrieved at mid latitudes;

    \item We find that to a first approximation, the opacity and pressure of the Aerosol-2 layer remains constant, although there are indications that both may be slightly reduced near 80$^\circ$S;

    \item We find that the retrieved opacity of the Aerosol-3 layer matches well the distribution of upper atmospheric haze seen at methane-absorbing wavelengths, and has highest abundance at the equator and $80^\circ$S, with secondary maxima at $30^\circ$S and $60^\circ$S;

    \item We find a smooth latitudinal distribution of `deep' mole fraction of methane (i.e., below the condensation level) in our simple `step' methane profile, with values of 6--7\% at the equator reducing to $\sim$3\% south of 25$^\circ$S;

    \item We find the South Polar Wave at $\sim$60$^\circ$S to be caused by a perturbation of the particles in the Aerosol-1 layer that reduces the single-scattering albedo at shorter wavelengths. The darkening appears identical to that which darkens the dark spot NDS-2018 and could be caused either by the addition of a chromophore in this layer that is darker than the background particles at all MUSE wavelengths, or by local heating at the $\sim$5-bar level that sublimates H$_2$S ice off to reveal the dark haze cores of the Aerosol-1 particles;

    \item We note a newly apprehended latitudinal variation in the reflectivity in narrow continuum wavelengths longer than 650 nm. We interpret these changes as being caused by variations in the scattering properties of the Aerosol-1 layer at wavelengths longer than 650 nm, with bright `zones' at the equator, 20$^\circ$S,  45$^\circ$S, and 75$^\circ$S, interspersed with darker `belts' at 5$^\circ$S,  30$^\circ$S, 50$^\circ$S, and 80$^\circ$S. This coloration could be caused by the addition of brighter, more scattering particles at the zone latitudes, or by local cooling at the $\sim$5-bar level that condenses more bright H$_2$S ice on to the Aerosol-1 particles. 

\end{enumerate}

The banding seen in the narrow windows of very low methane absorption longer than 650 nm is suggestive of a finer latitudinal scale variation than has been noted before on Neptune at the level of the H$_2$S/haze Aerosol-1 layer at $\sim$5 bar and indicates that the circulation of Neptune's atmosphere is even more complicated than that suggested by the `stacked circulation' models of \citeA{tollefson19} and \citeA{fletcher20}. Further work is needed to explore such changes more fully and recent observations with the James Webb Space Telescope should provide strong new constraints. In addition, the spatial deconvolution techniques developed for this study could be applied to existing ground-based IFU measurements from instruments such as Gemini/NIFS \cite{irwin11} and VLT/SINFONI \cite{irwin16} to extend the high spatial resolution analysis to longer wavelengths, and perhaps better constrain the latitudinal dependence of H$_2$S abundance. Further in the future, a space mission to one of the Ice Giants, comprising an orbiter and an entry probe would be invaluable in providing some measure of `ground truth' for aerosol models. In particular, it would be useful to observe with a probe the penetration depth of the UV flux to establish the pressure levels at which we may expect the photolysis of methane and other components. Meanwhile, an orbiter would provide reflectivity observations at higher phase angle, which will help to further constrain the allowable range of solutions that are consistent with observations.

\acknowledgments

We are grateful to the United Kingdom Science and Technology Facilities Council for funding this research (Irwin: ST/S000461/1, Teanby: ST/R000980/1). Glenn Orton was supported by funding to the Jet Propulsion Laboratory, California Institute of Technology, under a contract with the National Aeronautics and Space Administration (80NM0018D0004). Leigh Fletcher and Mike Roman were supported by a European Research Council Consolidator Grant (under the European Union's Horizon 2020 research and innovation programme, grant agreement No 723890) at the University of Leicester.  Santiago P\'{e}rez-Hoyos and Agustin S\'{a}nchez-Lavega are supported by the Spanish project PID2019-109467GB-I00 (MINECO/FEDER, UE), Elkartek21/87 KK- 2021/00061 and Grupos Gobierno Vasco IT-1742-22.

\section{Open Research}

The raw VLT/MUSE datasets studied in this paper (under ESO/VLT program: 0104.C-0187) are available from the ESO Portal at \url{https://archive.eso.org/eso/eso_archive_main.html}. The reduced raw and deconvolved `cubes' for the observation IDs: 6 -- 10, discussed in this paper, are available at \citeA{irwin23a}. Data files associated with this analysis are available at \citeA{irwin23b}.
The spectral fitting and retrievals were performed using the NEMESIS radiative transfer and retrieval algorithm \citeA{irwin08} and can be downloaded from \citeA{irwin22a} (or \url{https://github.com/nemesiscode/radtrancode}), with supporting website information at \citeA{irwin22b} (or \url{https://github.com/nemesiscode/nemesiscode.github.io}).

\bibliography{Neptune_Latitude}

%
%
%
%
%

\end{document}